Review Article

# Low-Dimensional Solid-State Single-Photon Emitters

Jinli Chen[1], Chaohan Cui[2,6], Ben Lawrie[3,4], Yongzhou Xue[5], Saikat Guha[2,6]*, Matt Eichenfield[2]*, Huan Zhao[3]*, Xiaodong Yan[1,2,5]*

1. Department of Materials Science and Engineering, University of Arizona, Tucson, AZ 85721, USA

2. James C. Wyant College of Optical Sciences, University of Arizona, Tucson, AZ 85721, USA

3. Center for Nanophase Materials Sciences, Oak Ridge National Laboratory, Oak Ridge, TN 37831, USA

4. Materials Sciences and Technology Division, Oak Ridge National Laboratory, Oak Ridge, TN 37831, USA

5. Department of Electrical and Computer Engineering, University of Arizona, Tucson, AZ 85721, USA

6. Department of Electrical and Computer Engineering, University of Maryland, College Park, MD 20742, USA

*corresponding author

Abstract:

Solid-state single-photon emitters (SPEs) are attracting significant attention as fundamental components in quantum computing, communication, and sensing. Low-dimensional materials-based SPEs (LD-SPEs) have drawn particular interest due to their high photon extraction efficiency, ease of integration with photonic circuits, and strong coupling with external fields. The accessible surfaces of LD materials allow for deterministic control over quantum light emission, while enhanced quantum confinement and light-matter interactions improve photon emissive properties. This review examines recent progress in LD-SPEs across four key materials: zero-dimensional (0D) semiconductor quantum dots, one-dimensional (1D) nanotubes, two-dimensional (2D) materials, including hexagonal boron nitride (hBN) and transition metal dichalcogenides (TMDCs). We explore their structural and photophysical properties, along with techniques such as spectral tuning and cavity coupling that enhance SPE performance. Finally, we address future challenges and suggest strategies for optimizing LD-SPEs for practical quantum applications.

# 1.Introduction

Photonic quantum technologies that harness the quantum properties of light (photons) to process quantum information have drawn increasing interest over the past decade. Single-photon sources [1–4] — isolated quantum systems designed to emit precisely one photon per excitation cycle — have garnered widespread research interest. These flying qubits play a crucial role in encoding, transmitting, and transducing quantum information, forming the foundation of many approaches to quantum computing[5], metrology[6], sensing[7], and secure communication[8].

Solid-state SPEs stand out by combining substantial control over optical properties with the scalability of solid-state systems[9–11], offering on-demand emission that is not available in other quantum light sources, such as those utilizing nonlinear optical techniques like spontaneous parametric down-conversion (SPDC)[12–14] and spontaneous four-wave mixing (SFWM)[15]. SPEs have been reported in various material platforms such as quantum dots (QDs)[16,17], rare earth ions[18–20], and defect centers[21] (e.g., nitrogen-vacancy (NV) centers in diamond[22–24]), offering high brightness, purity, and indistinguishability. Additionally, some SPEs exhibit optically addressable spin states, suitable for quantum sensing, transduction, and nuclear magnetic resonance spectroscopy[25].

SPEs in 3D bulk materials such as silicon[26,27] and diamond suffer from low (<10%) intrinsic photon extraction efficiency due to internal reflection, though extensive efforts have gone into patterning 3D photonic platforms with improved extraction efficiency[28]. In addition, integrating 3D SPEs into photonic structures[29] is usually not straightforward and requires demanding fabrication process[30–32]. LD-materials, on the other hand, offer high photon extraction efficiencies and comparably easy integration with photonic interfaces[33–35]. Their accessible surface and enhanced surface-volume ratio allows for deterministic SPE creation[36,37] and efficient tuning of emission properties[38,39]. Additionally, low-dimensional materials could host exotic physical phenomena, such as valley degrees of freedom[40] and exciton-magnon coupling[41,42], which pave the way for coherent control of photonic qubits via external fields[43,44].

This review examines the advancement in LD-SPEs developed from four key material systems: 0D semiconductor QDs, 1D nanotubes, 2D hBN, and 2D TMDCs (Fig. 1a). We begin by discussing the mechanism behind single photon emission (Sec. 2) and the fundamental properties of SPEs (Sec. 3). This is followed by an in-depth review of the structures, photophysical properties, state-of-the-art performance, applications, and challenges associated with LD-SPEs (Sec. 4-7). In Secs. 8 & 9, we explore techniques like spectral tuning and cavity coupling, which are widely utilized to enhance SPE performance. Finally, in Sec. 10, we summarize the key findings in LD-SPEs and provide perspectives on future research directions.

# 2.Single Photon Emitters: Realization and Excitation

Over the past four decades, SPEs have been successfully demonstrated in a wide variety of materials.

These include bulk materials such as diamond[45,46], silicon carbide (SiC)[47], silicon nitride (SiN)[48], aluminum nitride (AlN)[49,50] and gallium nitride (GaN)[51,52]; two-dimensional materials like hBN[53,54] and tungsten diselenide (WSe$_2$)[55–58]; one-dimensional materials like single-walled carbon nanotubes (SWCNTs)[59,60] and nanowires[61]; and zero-dimensional materials such as colloidal quantum dots[62], graphene quantum dots[33], and epitaxial quantum dots[63,64]. Single photon emission in these materials typically originates from two mechanisms: (1) color centers and (2) excitons confined within nanostructures or potential wells[34]. In both cases, a well-defined two-level energy structure is required to avoid unwanted emission from multiple optical transitions[65].

Color centers, usually observed in wide-bandgap systems such as diamond[31] and hBN[66], are defects in the crystal lattice where an atom is either missing or replaced by a different atom (i.e., vacancies or impurities). They disrupt the periodic potential within the solid, creating localized electronic states within the material's bandgap (Fig. 1b). These defects, which are usually found in nanocrystals, exfoliated layers, powders, and bulk materials, can be created through thermal annealing[67], femtosecond laser direct writing[68], electron or ion beam irradiation[69,70], mechanical indentation[71], and chemical/plasma treatment[72,73]. However, these techniques face physical limitations that hinder the consistent creation of identical emitters and frequently lead to the formation of unintended defects, thereby impeding the production of uniform SPEs.

Confined excitons, usually observed in low-dimensional materials such QDs[17], SWCNTs[74] and TMDCs[75], are bound states of excitons that are spatially localized (Fig. 1c). These regions, often caused by defects, strain, or size confinement, disrupt the uniform electronic potential, resulting in quantized energy states[76] that can be engineered through mechanical exfoliation[56], chemical vapor deposition (CVD)[77], chemical functionalization[78], and strain engineering[79]. However, these methods may result in spatial and spectral inhomogeneities due to intrinsic solid-state environment noise and fluctuations in fabrication conditions, complicating the generation of reproducible and indistinguishable SPEs. Thus, new techniques that can inherently ensure the production of identical SPEs are highly desired.

Excitation involves pumping electrons from the ground state to an excited state. We categorize the methods for exciting LD-SPEs based on various perspectives, including optical vs. electrical excitation, continuous wave (CW) vs. pulsed excitation, and resonant vs. non-resonant excitation, as detailed below.

**Optical vs. Electrical Excitation:** Optical excitation is widely used for SPEs and can be controlled by adjusting light intensity and wavelength. Jungwirth et al. (2016)[80] employed optical excitation to survey the temperature-dependent emission of point defects in multilayer hBN. They identified single-photon emission from individual SPEs by leveraging the spectral selectivity and high spatial resolution of optical excitation. Electrical excitation, alternatively, uses an electric current or voltage to stimulate SPEs, enabling single-photon emission through ambipolar emission (Fig. 1d), where both electrons and holes recombine, or unipolar emission (Fig. 1e), where a single type of carrier recombines with existing states. Clark et al.

(2016)[81] demonstrated electrically driven SPEs in WSe$_2$ and observed their electroluminescence (EL) intensities at cryogenic temperature, paving the way for on-chip SPEs in TMDCs. Fig. 2a & Fig. 2b illustrate the different mechanisms of optical and electrical excitations.

Cathodoluminescence (CL) microscopies have separately emerged as a powerful resource for near-field imaging of LD-SPEs[82–89]. CL microscopy leverages converged electron-beams in scanning (transmission) electron microscopes combined with far-field optical detection of photons generated by the material after electron-beam excitation in order to probe LD-SPEs with true nanoscale resolution well below the optical diffraction limit. While CL microscopies have been used to probe the photon statistics of SPEs, much of the literature exhibits photon bunching instead of photon antibunching under electron-beam excitation because the high-energy electron-beam can easily excite unwanted electronic transitions, resulting in the concurrent emission of photons from many excited states at once[83,84,90,91]. However, appropriate choices of electron-beam current do allow for nanoscale probes of photon antibunching[87,88,92]. CL microscopy is a particularly appealing tool for LD-SPEs because of the potential for in situ patterning and modification of SPEs and their environments. For instance, CL microscopy has been used for in situ monitoring of e-beam patterned defects in hBN[86,89], and e-beam induced etching in water vapor environments has been used to pattern nanoscale diamond cavities with *in situ* CL feedback[32]. Ultimately, the ability to image and pattern LD-SPEs at these length scales may lead to the development of new integrated quantum photonic systems with optimized cavity interactions designed to achieve ideal SPE properties.

**Continuous Wave (CW) vs. Pulsed Excitation**: CW excitation, which uses a steady energy source such as a laser or electric current, enables sustained single-photon emission but can reduce purity at higher excitation power. In contrast, pulsed excitation employs short, intense bursts of energy, offering advantages in time-resolved studies. Li et al. (2019)[93] utilized both CW and pulsed excitation to investigate the emission properties of hBN SPEs, reporting emission rates of 44 MHz under CW and 10 MHz under 80 MHz pulsed excitation. Notably, the purity of the SPEs under CW excitation reduced significantly as the excitation power increased, whereas purity under pulsed excitation remained high even at saturation power. Pulsed excitation also allows for time-gated correlation measurement that can improve measured single photon purity[94].

**Resonant vs. Non-Resonant Excitation:** Resonant excitation utilizes an excitation energy that exactly matches the optical bandgap[4]. This method allows for near-deterministic excitation of the emitter, minimizing excess energy that could cause unwanted emission or phonon-induced spectral broadening. However, resonant excitation requires complicated excitation or detection schemes, such as polarization filtering[95,96], phonon sideband (PSB) detection[97], and non-normal excitation[98], to separate the emitted photons from the excitation laser (Fig. 2c & Fig. 2d). Wang et al. (2019)[95] employed polarization filtering to collect single photons from SPEs in InGaAs quantum dot. They reduced the polarization loss to 3.8% instead of 50% by coupling to polarization-selective Purcell microcavities. Extra filtering requirements sometimes limit the polarization direction and intensity. Non-resonant excitation, on the other hand, uses energy higher than the bandgap (Fig. 2e). While simpler to implement, this method often results in lower-quality single photons and degrades the indistinguishability of the SPEs. Alternatively, 'quasi-resonant' excitation emerged to address the above issues[99]. In this approach, two near-resonant pulses are used to predictably excite the SPEs[100,101] (Fig. 2f), allowing for natural decay or stimulated decay with a second pulse[102]. Jayakumar et el. (2013)[99] utilized a two-photon excitation scheme on a InAs/GaAs QDs embedded into a microcavity. It allows for the deterministic generation of photon pairs, making the scheme suitable for generating time-bin entanglement.

## 3. Characterization of single-photon emission

SPEs are characterized by BPI values (Brightness, Purity, and Indistinguishability). Brightness (**B**) is quantified by PL intensity, which represents the number of collected photons per second. **B** is proportional to the excitation rate, quantum yield (QY) and collection efficiency. The QY quantifies the efficiency of photon emission in response to excitation, which is calculated from the number of emitted single photons at saturation normalized to the laser repetition rate. The measured value of **B** can vary depending on the measurement location: at the first collection element ($B_1$), coupled inside a single mode fiber or optical path ($B_2$), and at the detector ($B_3$)[11]. Fig. 3a shows a simple illustration of a fiber-coupled measurement scheme. **B** is wavelength dependent (Fig. 3b) and is a function of excitation power (inset Fig. 3b). The inset of Fig. 3b shows the PL intensity of a defect in hBN nanoflakes[93], which saturates at high power due to the finite availability of excited states and competition with non-radiative recombination processes, such as Auger recombination.

Single-photon purity (**P**) is quantified by the second order correlation function $g^{(2)}(\tau)$, which is given by

$$g^{(2)}(\tau) = \frac{\langle n_1(t) n_2(t+\tau) \rangle}{\langle n_1(t) \rangle \langle n_2(t+\tau) \rangle},$$

where $n_i(t)$ is the number of counts registered by the detector $i$ at time $t$, $\tau$ is delay time between photon detection events in two detectors, and the $<...>$ is the time average operator[103]. $g^{(2)}(\tau)$ measures the probability that the source produces at most one photon per excitation event. Sometimes different forms of $g^{(2)}(\tau)$, such as $1 - g^{(2)}(0)$ or the Mandel Q function[104] $Q(\tau) = g^{(2)}(\tau) - 1$ may be used to quantify **P**. Fig. 3c shows a simple experimental Hanbury–Brown and Twiss (HBT) setup[105], which can be used to approximate $g^{(2)}(\tau)$ in the limit where the probability of measuring more than one photon on one detector at a time is much smaller than the probability of a single photon detection event. Photons from a light source are split by a 50:50 beam splitter and directed to two detectors, and the time correlation between detection events is measured to assess the statistical properties of the photon stream. An ideal n-photon Fock state exhibits $g^{(2)}(0) = 1 - \frac{1}{n}$, and an ideal single photon emitter therefore exhibits $g^{(2)}(0) = 0$ with $g^{(2)}(0) < g^{(2)}(\tau) \, \forall \, \tau$. When more than one emitter with identical brightness is excited simultaneously, $g^{(2)}(0) > 0.5$, though $0 < g^{(2)}(0) < 0.5$ is often observed when a single SPE is excited together with other weak SPEs and background fluorescence from the host material.

Fig. 3d shows $g^{(2)}(\tau)$ for a 36 nm thick GaSe crystal under CW excitation[79], characterized by an exponential dip in coincidence counts at zero-time delay. This antibunching dip is fitted by $g^{(2)}(\tau) = A exp(-\frac{\tau}{\tau_0})$, where $\tau_0$ is a time constant determined by the emission lifetime $\tau_e$ and pumping time $\tau_p$ following $\frac{1}{\tau_0} = \frac{1}{\tau_e} + \frac{1}{\tau_p}$. The inset of Fig. 3d shows $g^{(2)}(\tau)$ of color centers in an hBN flake[106] coupled to a microcavity under pulsed excitation, with a series of peaks separated by the reciprocal of the excitation laser repetition rate, and suppressed coincidences at zero-time delay.

Indistinguishability (***I***) is quantified by $V_{HOM}$, the visibility in the Hong-Ou-Mandel (HOM) experiment[107,108]. Fig. 3e illustrates a HOM experiment, where single photons are split into two paths by the first beam splitter with a delay time $\tau$ introduced between the paths. At the second beam splitter, quantum interference occurs, and if the photons are indistinguishable, they will exit together, leading to reduced coincidence counts for $\tau = 0$. By polarization filtering or adjusting $\tau$, we can obtain $V_{HOM}$[108], given by $V_{HOM} = \text{Tr}(\widehat{\rho_1} \, \widehat{\rho_2}) = \frac{\text{Tr}(\widehat{\rho_1^2}) + \text{Tr}(\widehat{\rho_2^2}) - O(\widehat{\rho_1}, \widehat{\rho_2})}{2}$, where $O(\widehat{\rho_1}, \widehat{\rho_2}) = |\widehat{\rho_1} - \widehat{\rho_2}|^2$ is the operational distance between the states of the two photons $\widehat{\rho_1}$ and $\widehat{\rho_2}$. The ***I*** is related to the optical coherence time $T_2$ and spontaneous emission time $T_1$ of SPEs by the approximate relation $I = T_2/2T_1$. The pure dephasing rate $\gamma$ represents the rate at which a quantum system loses its coherence due to environmental interactions without energy dissipation, following $\frac{1}{T_2} = \frac{1}{2T_1} + \gamma$. For ideal SPEs without dephasing, $T_2/2T_1 = 1$. In practice, this ratio is smaller than one because $T_2$ is suppressed by interactions between SPEs and their environment, such as charge and spin noise[109] and phonon scattering[110]. $T_2$ can be increased by materials engineering, while $T_1$ can be reduced by Purcell enhancement achieved through local control of electromagnetic fields. Utzat et al.[111] demonstrated that halide perovskite quantum dots (PQDs) $CsPbBr_3$ exhibit efficient single-photon emission with $T_2 \approx 80$ ps and $T_1 \approx 210$ ps at 4 K, making PQDs attractive in realizing high ***I*** SPEs among colloidal quantum dots (CQDs). Fig. 3f shows the $g^{(2)}_{HOM}(\tau)$ of nanotube defects (NTDs) in $sp^3$-functionalized SWCNTs coupled with an optical cavity[112], leading to a $V_{HOM}$ up to 0.65 and a 217-fold enhancement in visibility.

The BPI values influence the design rules for SPEs. Table. 1 demonstrates the performance of SPEs in quantum applications, where source efficiency refers to the probability of collecting a photon in each excitation pulse (proportional to ***B***). In QKD protocols, such as BB84[113], both ***P*** and ***B*** contribute to improving the secure key rate. In other applications, such as Greenberger–Horne–Zeilinger (GHZ) state generation[114–116], all BPI values are crucial for enhancing fidelities and overall efficiencies. Other factors also affect the design rules of SPEs. For example, the emission wavelength determines the transmission properties in various media[17] (Fig. 1a). SPEs with telecommunication band emission minimize the transmission losses in optical fibers, enable long-distance QKD[117] and are promising for quantum internet construction. Emission in the ultraviolet range is ideal for free-space transmission. SPEs with electrical excitation capability are ideal for integration into Complementary Metal-Oxide-Semiconductor (CMOS) circuits[34,118]. SPEs that can emit photons at room temperature or higher[119] reduce the energy cost associated with cooling systems. SPEs that emit in a Gaussian mode facilitate seamless waveguide coupling[120].

In Sec. 4-7, we review the progress of LD-SPEs categorized by their materials and emission wavelengths (Fig. 1a). We discuss (1) semiconductor QD SPEs with emission wavelengths spanning approximately from 280 nm to 1550 nm, (2) nanotube SPEs from 570 nm to 2000 nm, (3) hBN SPEs from 300 nm to 850 nm, and (4) TMDCs from 600 nm to 1550 nm. We focus in particular on SPEs capable of emitting single photons at telecommunication wavelengths, based on materials such as InAs/InP QDs, SWCNTs, and MoTe$_2$.

## 4. Semiconductor Quantum Dots

LD-SPEs have been demonstrated in CQDs[121], graphene QDs (GQDs)[33,122], and epitaxially grown QDs (EQDs)[17]. Single-photon emission in QDs originates from excitons formed within discrete energy levels due to quantum confinement.

CQDs are semiconductor nanocrystals with core sizes typically ranging from 2 to 10 nanometers, synthesized in a colloidal solution[121]. CQD SPEs are attractive due to their flexibility in synthesis[123], ease of integration, and ability to operate at room temperature[124]. CQD SPEs offer tunable emission wavelengths, which can be controlled by adjusting their size, morphology and structure[121,125]. Krishnamurthy et al. (2021)[126] demonstrated PbS/CdS SPEs with tunable emission wavelengths covering the telecom S band (1460–1530 nm) and O band (1260–1360 nm) at room temperature. The tunability was achieved by adjusting the core size and shell thickness of PbS/CdS CQDs. Chandrasekaran et al. (2017)[124] demonstrated near-blinking free, high purity ($g^{(2)}(0) = 0.03$) InP/ZnSe core/shell QD SPEs at 629 nm by leveraging tris(diethylamino)phosphine as the phosphorus precursor. The SPEs exhibited sharp room-temperature spectra and photostability. CQD SPEs in general face challenges such as low PL stability, often manifesting as blinking or intermittent fluorescence, due to strong Auger recombination.

As a specific type of CQD, Halide perovskites QDs (PQDs) exhibit room-temperature single photon emission, near-unity QY and high photostability. PQDs are synthesized through methods[127] such as hot injection[128,129], ligand-assisted precipitation[130,131], and ultrasonic synthesis[132,133]. Liu et al. [134] demonstrated 100% QY in CsPbI$_3$ PQDs at room temperature, employing a synthetic protocol involving the introduction of an organometallic compound, trioctylphosphine- $PbI_2$, as the reactive precursor. Tang et al.[135] demonstrated $CsPbBr_3$/CdS core/shell PQDs with non-blinking PL and a high QY of 90%, attributed to the reduction of electronic traps within the stable core/shell structure. Utzat et al.[111] showed that $CsPbBr_3$ SPEs exhibit fast emission lifetimes of 210-280ps (Fig. 4a) with a large $T_2/2T_1$ ratio (~0.2) and stable emission over several minutes at cryogenic temperatures at 520nm (2.38eV). Zhu et al. [136] reported $CsPbI_3$ SPEs with 98% $\boldsymbol{P}$ ($g^{(2)}(0) = 0.02$) at room temperature, demonstrating that enhanced quantum confinement is key to improving purity by suppressing biexciton emission. Kaplan et al.[137] measured the $g_{HOM}^{(2)}(\tau)$ of single photons from $CsPbBr_3$ CQDs at 3.9 K, showing corrected visibilities up to $0.56 \pm 0.12$ without using any cavity. PQDs exhibit tunable wavelengths ranging from 400 nm to 800 nm that can be controlled by adjusting their composition, size, morphology, and dimensions. Jun et al. [138] coupled $CsPbBr_3$ nanocrystals to circular Bragg grating cavities, exhibiting 5.4-fold PL enhancement with lifetimes reduced to less than 100 ps. $CsPbI_3$ PQDs integrated into optical microcavities have also been shown to exhibit narrow room-temperature linewidths (~1 nm).[139]

CQD SPEs face several bottlenecks waiting to be addressed: strong Auger recombination reduces the PL intensity; intense multiexciton emission restricts HOM experiments to cryogenic temperatures; blinking leads to poor photostability at ambient conditions; and there are no reports to date of telecommunications band operation.

Graphene QDs (GQDs) are atomically thin fragments of graphene, typically consisting of 1 or 2 layers with lateral sizes below 10 nm[140]. Compared to graphene, GQDs exhibit desirable properties for SPEs, such as a bandgap opened as a result of quantum confinement and tunable physical properties enabled by geometry engineering and chemical functionalization. GQD SPEs are synthesized by bottom-up methods, such as molecular fusion, allowing control over size, morphology, doping, functionalization, and synthesis techniques. Zhao at el.[33] demonstrated GQD SPEs at room temperature with high $\boldsymbol{B}$, $\boldsymbol{P}$ ($g^{(2)}(0) = 0.05$) and no blinking (Fig. 4b). The emission wavelengths were tunable from ~650nm to ~750nm by chlorine functionalization. Theoretical studies indicate that amino-group-functionalized GQDs exhibit single-photon emission.[141] GQD SPEs commonly suffer from low $\boldsymbol{I}$[141]. Coupling GQD SPEs to photonic cavities remains an outstanding challenge that will be essential to the development of GQDs as technologically viable SPEs.

EQDs are synthesized through layer-by-layer techniques such as Molecular Beam Epitaxy (MBE) and Metal-Organic Chemical Vapor Deposition (MOCVD)[142]. SPEs with tailored nanostructures can be formed by tuning growth conditions, strain, and utilizing pre-patterned substrates along with post-growth techniques like etching and lithography[143,144]. EQD SPEs cover a broad emission wavelength range, from the ultraviolet (below 280 nm) to the telecom band (around 1550 nm) (Fig. 4c). Emission in the UV range is achieved using III-nitrides, particularly GaN/AlGaN EQDs. Holmes et al.[16] demonstrated that

GaN/AlGaN SPEs maintain high $P$ ($g^{(2)}(0) = 0.34$) at temperatures up to 350K. Fig. 4d illustrates the variation of PL intensity with temperature. The robustness is attributed to the large biexciton binding energies (~50 meV)[145].

InGaN, InGaN/GaN, InP, and various II–VI EQDs[146,147], such as CdSe/ZnSe have been used to realize SPEs with emission in the visible region. Fedorych et al.[147] demonstrated CdSe/ZnSSe/MgS EQD SPEs with $g^{(2)}(0) = 0.16$ under CW excitation at 300K. K. Quitsch et al.[148] demonstrated electrical excitation in the same structure. Deshpande et al.[149,150] demonstrated InGaN/GaN SPEs operating at 620 nm with $g^{(2)}(0) = 0.37$ under pulsed excitation and $g^{(2)}(0) = 0.32$ under CW excitation at 280 K. Cho et al.[151] realized InGaN SPEs operating at 473 nm with $g^{(2)}(0) = 0.11$ at 10K.

III-arsenide materials, such as InAs/GaAs EQDs, have been used to achieve single photon emission in the near-infrared (NIR) region with near-unity $I$. He et al.[152] demonstrated InAs/GaAs SPEs operating at a wavelength of 940 nm with $g^{(2)}(0) = 0.012$ and ~97% HOM visibility at 4.2K. The combined purity and indistinguishability of this source allowed for the realization of a quantum CNOT gate suitable for generating entangled states from the SPE. Somaschi et al.[153] demonstrated InGaAs SPEs operating at a wavelength of 890 nm with $g^{(2)}(0) = 0.0028$ and 99.56% visibility under resonant excitation at 4.2K (Fig. 4e), along with $B$ more than an order of magnitude higher than SPDC sources. Wang et al.[154] demonstrated 3-,4-, and 5- photon boson sampling with sampling rates of 4.96 kHz, 151 Hz and 4 Hz respectively. This application leveraged the high $P$ ($g^{(2)}(0) = 0.027$) and $I$ (0.900) of InAs/GaAs EQD SPEs integrated with a micropillar cavity and low-loss photonic circuits. Schweickert et al. (2018)[155] achieved $g^{(2)}(0) = (7.5 \pm 1.6) x 10^{-5}$ at a wavelength of 790 nm by utilizing two-photon excitation of the biexciton state in a GaAs/AlGaAs quantum dot at 4 K.

The telecommunication band is primarily covered by InAs/InP and InAs/GaAs EQDs due to their narrow bandgaps (0.2eV~1.2eV). The emission wavelengths of InAs/InP SPEs can be tuned to 1550 nm (telecom C-band)[156]. Takemoto et al.[117] demonstrated InAs/InP SPEs with high purities ($g^{(2)}(0) = 0.002$ after background correction) by employing an optical horn structure[157] to enhance photon extraction efficiency and demonstrated QKD over 120 km. Miyazawa et al.[158] demonstrated InAs/InP SPEs operating at 1500 nm with $g^{(2)}(0) = 4.4 x 10^{-4}$ at 8 K (Fig. 4f). Muller et al.[159] showed that InAs/GaAs EQD SPEs exhibit antibunching at 1550 nm with $g^{(2)}(0) = 0.11$ at 4 K, and they leveraged a biexciton cascade mechanism to generate entangled photons with entanglement fidelity of 87% that remained stable at temperatures up to 94 K. Nawrath et al.[160] demonstrated InAs/InGaAs/GaAs SPEs operating at 1550 nm with $g^{(2)}(0) = 0.072$ under near-resonant excitation at 4K.

EQD SPEs demonstrated room-temperature emission in the UV or visible range, with tunable emission wavelengths controlled by adjusting the material's stoichiometry during growth[17]. Thanks to the maturity of EQD growth techniques, these SPEs exhibit promising scalability for quantum applications such as boson sampling[154,161,162], linear cluster state generation[163–165], QKD[166,117], quantum logic gate operation[167] and quantum teleportation[168,169]. While promising, EQD SPEs face several challenges: HOM measurements are only favorable at cryogenic temperatures due to strong dephasing rates at room temperature; it remains challenging to achieve usable photon indistinguishability for photons emitted from distinct EQD SPEs, and the difficulty in achieving electrical excitation limits the potential for integration into on-chip devices.

## 5. Nanotubes

1D nanotubes SPEs are primarily realized using two materials: boron nitride nanotubes (BNNTs) and SWCNTs. BNNTs, structurally similar to rolled boron nitride sheets, are known for their wide bandgap and high thermal and chemical stability[170]. BNNT SPEs exhibit room-temperature single-photon emission in the 570–610 nm range[171,172]. Ahn et al.[171] studied point defects in 50-nm-diameter BNNTs, demonstrating $g^{(2)}(0) = 0.38$ under non-resonant excitation at room temperature, with spectral modulation enabled via a NIR control laser. Gao et al.[172] observed spin defects in BNNTs with a spin S = 1/2 ground state and no intrinsic quantization axis, in BNNTs, exhibiting $g^{(2)}(0) < 0.5$ under non-resonant excitation at room temperature. These spin defects have potential for magnetic sensing, with a typical DC magnetic field sensitivity of $\sim 80$ µT$/\sqrt{Hz}$.

SWCNTs consist of covalently bonded carbon atoms arranged in an ordered tubular structure, with their diameter and roll-up angle defined by the chiral index (**n**, **m**) where **n** and **m** specify the wrapping direction of the graphene lattice. The chiral index also defines the emission wavelength for intrinsic SWCNTs. SWCNTs stand out as candidates for SPEs due to their structure-specific NIR PL[173] (Fig. 5a). Single-photon emission from SWCNTs originates from excitons confined in potential wells created through non-covalent or covalent functionalization.

Non-covalent functionalization creates localized potential wells through unintentional molecular adsorption or local inhomogeneities at the interface with the surrounding matrix or substrate[174], allowing control of exciton diffusion and inducing strong photon antibunching. This approach leverages the sensitivity of SWCNTs to their dielectric environment while preserving their excellent optical characteristics.[175] Högele et al.[176] demonstrated strong photon antibunching in CoMoCat SWCNTs encapsulated in sodium dodecylbenzenesulfonate, achieving $g^{(2)}(0) = 0.03$ at 10 K. The SPEs maintained high **P** while tuning the emission wavelength between 855 and 885 nm by adjusting the temperature from 4.2 K to 25 K. Khasminskaya et al.[177] demonstrated electrically excited SWCNT SPEs at 1.6K by integrating HiPco SWCNTs with a waveguide circuit. The **P** ($g^{(2)}(0) = 0.49$) of these SPEs was limited by the timing resolution of the detector (Fig. 5b). Raynaud et al.[174] quantified potential energy disorder along the tube axis using hyperspectral imaging and quasi-resonant excitation spectroscopy, revealing that interface roughness leads to exciton localization at low temperatures, resulting in segmented photoluminescence lines and random potential traps with a 70 nm spacing and 20 meV energy spread. However, non-covalent functionalization persists its weak interaction with SWCNTs only at cryogenic temperature, making the fabricated SPEs unsuitable for room-temperature applications.

Covalent functionalization has enabled a variety of approaches for creating localized excitons through methods such as oxygen doping, diazonium salts, DNA, or photoexcited aromatics-based functionalization. Ma et al.[59] demonstrated solitary oxygen dopant SWCNT SPEs with $g^{(2)}(0) = 0.32$ at 298K in the $1100 - 1300$ nm wavelength range, achieved by incorporating undoped (6,5) SWCNTs into a $SiO_2$ matrix. Zheng et al. (2021)[178] reported single-photon emission from coupled defect states in DNA-functionalized SWCNTs, with $g^{(2)}(0) = 0.27$ at room temperature, where guanine defects in ssDNA strands created multiple coupled trapping sites due to dense covalent functionalization.

Aryl $sp^3$ defects created through diazonium-based reactions have emerged as promising candidates for SPEs, offering stable, shot-noise limited emission. These defects are synthetically tunable, allowing for enhanced trapping potentials and red-shifted emission, particularly in large-diameter tubes emitting at telecom wavelengths. He et al.[74] demonstrated stable SPEs with $g^{(2)}(0) = 0.01$ and telecom wavelength (1550 nm) emission from SWCNT $sp^3$ defects at room temperature. They explored the room-temperature PL of SWCNTs with different chiral indices, functionalized with Cl$_2$-Dz and MeO-Dz. By using DOC and PFO-bpy coatings, they achieved emission in the 1300–1550 nm range for (6,5), (7,5), and (10,3) SWCNTs (Fig. 5c). While aryl $sp^3$ defect SPEs typically exhibit low $T_2/2T_1$ ratios (around 1%), this can be improved by coupling to optical cavities[173]. Husel et al.[112] conducted HOM experiments on individual NTDs coupled to an optical microcavity (Fig. 5d), achieving a visibility of 0.65 and $g^{(2)}(0) = 0.31$ at room temperature in the telecom band. This was the first room-temperature demonstration of cavity-enhanced $I$ for LD-SPEs, despite the high dephasing rate (Fig. 5e). Unfortunately, covalently functionalized SWCNT SPEs suffer from low QY (10–30%) and are prone to spectral diffusion and blinking[173]. Further, the $B$ of SWCNT SPEs is limited by strong exciton-exciton annihilation (EEA) arising from their one-dimensional nature. These limitations hinder the use of SWCNT SPEs in quantum applications. Advances in defect engineering and cavity coupling offer potential solutions to these challenges.

## 6. Few-Layer Hexagonal Boron Nitride

Two-dimensional hBN is a wide-bandgap insulator (E$_g$~5.97 eV) with a graphene-like honeycomb lattice of alternating boron and nitrogen atoms. The wide bandgap makes hBN SPEs robust against thermal fluctuations, enabling stable single-photon emission at room-temperature. Few-layer hBN SPEs show promising characteristics such as high BPI values and spin-photon interfaces. hBN SPEs have been successfully integrated with photonic circuits in initial demonstrations of scalable quantum photonic technologies[38,39]. Single photon emission in hBN arises from trapped excitons at defect sites, including nitrogen vacancies ($V_N$), boron vacancies ($V_B$), anti-site carbon vacancies ($V_N C_B$), and anti-site nitrogen vacancies ($V_N N_B$)[179] (Fig. 6a), which can be introduced by annealing[67,180], electron beam[181–183] or ion beam[69,182] irradiation, nanopillars[184,185], plasma processing[72,73] or femtosecond pulses[186,68]. Theoretical studies[187–190] have proposed new type of defects, such as $C_2 C_N$ and $C_2 C_B$ carbon clusters[191,192], as candidates for SPEs in hBN. hBN SPEs exhibit zero phonon lines (ZPLs) across the NIR-visible range ($\sim$560 –780 nm) [67,80] and the UV range ($\sim$300 nm) [87,193].

Tran et al.[53,54] first demonstrated single-photon emission from monolayer and multilayer hBN SPEs, with stable emission at 623 nm over 10 min and $g^{(2)}(0) < 0.5$. Fig. 6b illustrates the optical characterization method they used, where excitation and emission are separated by a dichroic mirror during measurements. Bourrellier et al.[87] demonstrated UV hBN SPEs using electron beam irradiation, achieving operation at 300 nm with postprocessed $g^{(2)}(0) = 0.2$ at room temperature. Previously, this emission wavelength was only achieved by III-nitride EQD SPEs.

Compared with many other platforms, hBN SPEs show reproducible emission wavelengths. Fournier et al.[183] demonstrated hBN SPEs operating at 436 nm that were created at controlled locations using electron beam irradiation. The local irradiation process activates SPE ensembles with submicron precision, leading to ZPLs consistently centered at 436 ± 1 nm (Fig. 6c). Horder et al.[194] employed resonant excitation to characterize the emission line shape, demonstrating coherence of optical transitions through the observation of Rabi oscillations. Fournier et al.[195] conducted HOM experiments on the blue hBN SPEs, demonstrating corrected HOM visibility of 0.56 ± 0.11 at cryogenic temperatures (Fig. 6d). The blue hBN SPEs exhibit spectral stability, room-temperature operation, ultra-narrow linewidth, and high *I* under non-resonant excitation, making them favorable for quantum frequency conversion (QFC) to telecommunications wavelengths.

Li et al.[93] demonstrated hBN SPEs coupled to metal-dielectric antennas, achieving near-unity light collection efficiency (98%) and a QY of 12% (Fig. 6e). The SPEs exhibited high *B* ($10^7$ cps at $10^1$ mW excitation power) and maintained single-photon emission under excitation powers up to 8mW at room temperature. Vogl et al. (2021)[106] demonstrated hBN SPEs integrated with a tunable microcavity[196] consisting of a hemispherical and flat mirror, achieving corrected $g^{(2)}(0) = 0.0064$ at room temperature (Fig. 6f). They measured single-photon interference with Michelson-type interferometers, demonstrating interferometric visibilities of up to 98.58%.

These properties make hBN SPEs well-suited for several quantum applications. White et al.[197] demonstrated quantum random number generation (QRNG) using hBN SPEs coupled to an on-chip photonic waveguide structure at room temperature. Samaner et al.[198] integrated an hBN SPE into a QKD system based on the B92 protocol, achieving a sifted key rate of 238 bps with a quantum bit error rate of

8.95% at a 1 MHz clock rate. Scognamiglio et al.[199] demonstrated that hBN SPEs operating at 417 nm show promise for underwater quantum communications.

One of hBN SPE's most unique properties is its compatibility with spin-based quantum sensing[200]. Optically active spin defects, such as nitrogen vacancies[201,202], exhibit a ground-state spin that can be optically addressed and manipulated. The manipulation of the spin states through external magnetic fields[203], temperature variation[204,205], or strain[206,204] forms the basis for their application in quantum sensing[207,208]. These defects have the potential to detect minute changes in environmental parameters like magnetic fields or temperature at the nanoscale, leading to applications in high-sensitivity quantum sensing devices (Fig. 7a). One of the most powerful techniques for characterizing and manipulating the spin properties of defects in quantum materials is Optically Detected Magnetic Resonance (ODMR)[209]. ODMR enables the detection of spin transitions in a defect's electronic structure by combining microwave excitation and optical readout (Fig. 7b). In the case of hBN spin defects, this method involves monitoring the fluorescence from the defect centers while applying microwave radiation to induce spin transitions between different energy levels.

The first spin signature for hBN defects came from ensembles attributed to a boron vacancy ($V_B^-$) with a broad optical emission spectrum centered at ~800 nm[203,206]. These defects, which exhibit a broad optical emission spectrum centered around 800 nm, are characterized by a ground-state spin triplet (S = 1). The spin axis aligns along the c-axis of the hBN crystal, with a zero-field splitting of D $\approx$ 3.45 GHz separating the spin $|0>$ and spin $|\pm1>$ sublevels[204,210] (Fig. 7c). Through careful laser pulsing, the $V_B^-$ defect's spin states can be initialized, manipulated, and optically read out. However, due to their relatively low optical quantum efficiency, ODMR measurements have been limited to ensemble-level experiments rather than single-spin resolution.

In contrast, carbon-related spin defects in hBN[211,212], some also with a spin triplet (S = 1), exhibit advantages for single-photon emission (Fig. 7d). These defects, which emit light in the visible spectrum range, demonstrate bright room-temperature emission with more than 80% of the emission occurring from the ZPL[213,214]. They also show high ODMR contrast over 30% (Fig. 7e) and a long dephasing time, exceeding 100 ns[215,216]. These features make them highly suitable for practical quantum sensing applications. Notably, these carbon-based defects achieve an estimated DC magnetic field sensitivity of approximately 3 μT /$\sqrt{Hz}$,[217] placing them on par with the well-established NV centers in diamond, which are commonly used for magnetometry. Interestingly, some carbon-related defects show no zero-field splitting in their ODMR curves, indicating that these are spin doublets (S = 1/2)[212,218,219]. The coexistence of spin triplet and spin doublet states within a single hBN crystal matrix adds another layer of versatility to the material, making it an attractive platform for quantum sensing across a range of applications. The ability to integrate different types of spin states within a single host material opens the door to more flexible and tunable quantum devices.

Optically active spin defects in hBN bring two primary advantages over traditional NV centers in diamond, particularly for quantum sensing applications. First, quantum sensors in hBN have higher photon extraction efficiency: The 2D nature of hBN provides superior photon extraction efficiency compared to diamond. In hBN, the emission originates from the surface, minimizing internal photon losses caused by total internal reflection—a problem commonly encountered in bulk diamond. This enhanced efficiency can improve the signal-to-noise ratio in quantum measurements, making hBN a promising alternative for quantum sensing. Second, the quantum sensors in hBN could potentially offer better sensitivity and spatial resolution. Due to its atomically thin, planar structure, hBN quantum sensors can be positioned just a few ångströms away from the target object, leading to unprecedented sensitivity and spatial resolution. This capability is particularly important for applications such as imaging magnetic domains in 2D materials, where proximity to the sample is critical. The atomically smooth surface of hBN further enhances the sensing potential by reducing signal interference from surface roughness. The unique properties of hBN defects have already been demonstrated in experimental setups, including quantum microscopes[220,221] and fiber-integrated devices[222] (Fig. 7f). As research progresses, hBN is likely to play a pivotal role in the future of quantum technology, providing a versatile, high-performance platform for both fundamental studies and practical quantum devices.

Several challenges hinder the broader application of hBN SPEs. The uniformity of hBN SPEs is limited due to inherent challenges in material manipulation and control at the nanoscale that prevent the consistent creation of identical defects and often result in the formation of unintended defect types. Most hBN SPE ZPLs exist at wavelengths of 400-800 nm, with telecom wavelength emission yet to be demonstrated, limiting integration into optical fiber applications. Both the QY (current record of 12%)[93] and $I$ (current record of 56%)[195] of hBN SPEs must be improved to realize technologically relevant hBN-based quantum photonic applications.

## 7. Few-Layer Transition Metal Dichalcogenides

The TMDCs have a layered $MX_2$ structure with hexagonally arranged X-M-X units[223] (Fig. 8a). The 2D TMDCs offer strong light-matter interaction[224,225], direct bandgaps, large exciton binding energies (0.5-1 eV)[226,227] and valley degrees of freedom[228], allowing for circularly polarized excitonic optical transitions and efficient tuning via magnetic field. The 2D TMDC-based SPEs offer high photon extraction efficiency, ease of coupling with external fields, and seamless integration into photonic circuits[38,39]. Single-photon emission in TMDCs originates from excitons trapped by localized strain[229] or defects[230]. Strain is introduced using bubbles[231,232], patterned nanostructures[233,234], nanopillars[75,235] or atomic force microscopy (AFM) tips[236,237] to funnel excitons into localized regions (Fig. 8b). Defects are introduced by material growth[55–58], electron beam[238,239] and ion beam irradiation[232].

Over the past decade, SPEs have been demonstrated in TMDCs such as $WSe_2$[55–58], $WS_2$[233,240], $MoS_2$[241–

[243], MoSe$_2$[244,245] and MoTe$_2$[38]. Palacios-Berraquero et al.[240] demonstrated electrically excited SPEs in layered WSe$_2$ and WS$_2$. The SPEs in WSe$_2$ emitted at 760nm with $g^{(2)}(0) = 0.29$, while the SPEs in WS$_2$ emitted at 640nm with $g^{(2)}(0) = 0.31$. The SPEs demonstrated seamless integration into the electrical excitation device, which was based on a single tunnelling heterojunction design. They further demonstrated arrays of WS$_2$ and WSe$_2$ SPEs by using nanopatterned silica substrates[233]. The SPEs showed visible-range PL, with WSe$_2$ emitting in the 730-820 nm range and WS$_2$ typically in the 610–680 nm range. The SPEs also showed PL stability, with spectral wandering remaining below 0.5 meV over a period of 1–2 minutes. Klein et al.[241] patterned SPEs in monolayer MoS$_2$ with helium ion irradiation and achieved emission at 705nm and $g^{(2)}(0)$=0.23 at cryogenic temperatures. Yu et al.[245] demonstrated MoSe$_2$ SPEs operating at 785nm with $g^{(2)}(0) = 0.29$ at cryogenic temperatures. Zhao et al.[38] demonstrated MoTe$_2$ SPEs spanning the telecom region from 1080 nm to 1550 nm with $g^{(2)}(0) = 0.058$ under pulsed excitation and $g^{(2)}(0) = 0.181$ under CW excitation. Significant efforts have been made to raise the operating temperature of TMDC-based SPEs beyond the cryogenic range. Parto et al.[239] used a combination of strain engineering via nanoscale stressors and defect engineering via electron-beam irradiation to pattern WSe$_2$ SPEs with $g^{(2)}(0) = 0.05$ at 5K, and $g^{(2)}(0) = 0.27$ at 150K.

The weak van der Waals (vdW) interactions between TMDC layers allow for incommensurate stacking, which can result from relative rotation between the layers or, in heterobilayers, from lattice mismatch. The interplay between lattice mismatch and interlayer electronic coupling leads to the formation of interlayer excitons (IXs), where the exciton's wavefunction spans both layers. For type-II band alignment, IXs possess out-of-plane electric dipole moments, enabling Stark tuning over a broad spectral range. Compared to intralayer excitons, IXs exhibit extended radiative lifetimes of order hundreds of nanoseconds and microsecond-scale valley lifetimes due to reduced electron-hole overlap. Combined with localized strain fields and in-gap defect states, such IXs lead to single-photon emission. Zhao et al.[246] demonstrated SPEs based on Γ- defect IXs in MoS$_2$/WSe$_2$ heterobilayers and heterotrilayers using nanopillars with a gold substrate. The nanopillars were used to create strain fields and defects, while the gold substrate quenched PL from the homogeneous region. The SPEs emitted at 855–1078 nm with $g^{(2)}(0) = 0.01$ under pulsed excitation at 10K. When the lattice mismatch and/or interlayer twist between the TMDC bilayers is small, a moiré lattice is formed with lattice constants of hundreds even thousands of times larger than the sublattice. This moiré lattice generates a periodic variation in the potential energy landscape, known as the moiré potential, ranging from a few millielectronvolts to tens of millielectronvolts[247]. Such a moiré potential can trap excitons, creating an array of moiré excitons that each serve as SPEs. Seyler et al.[248] demonstrated Moiré potential-trapped valley IXs in MoSe$_2$/WSe$_2$ heterobilayers, confirmed by PL and the g-factor measurements of the emitters (Fig. 8c). Baek et al.[249] realized MoSe$_2$/WSe$_2$ heterobilayer SPEs operating at 885nm with $g^{(2)}(0) = 0.28$. As IXs possess vertical dipoles, they demonstrated that the photon energy could be tuned (~40meV) via DC stark effect. Chuang et al.[250] demonstrated an improvement in *P* from 60% to 92% for WSe$_2$ SPEs by adding a graphene cap layer that quenched the background PL through fast interlayer charge transfer, preventing radiative recombination via long-lived defect-bound exciton states.

Substantial work has gone into cavity enhancement of TMDC SPEs. For instance, Luo et al.[235] demonstrated WSe$_2$ SPEs coupled to plasmonic cavities, achieving a Purcell factor up to 551 and an enhanced QY of up to 65% (average 44%) (Fig. 8d). Sortino et al.[251] coupled WSe$_2$ SPEs to GaP dielectric nano-antennas, demonstrating a $10^2$ to $10^4$ enhancement of the PL intensity. Von Helversen et al. (2023)[252] demonstrated WSe$_2$ SPEs with $g^{(2)}(0) = 0.036$ under pulsed excitation at cryogenic temperatures (Fig. 8e). Unfortunately, many TMDC SPEs exhibit low *I* due to their high dephasing rate. Drawer et al.[253] demonstrated WSe$_2$ SPEs with $g^{(2)}(0)$=0.047 and 2% visibility by coupling to a tunable optical cavity (Fig. 8f), though it is not clear at this stage how much improvement on that visibility can be expected in optimized photonic platforms.

There is now a limited literature describing TMDCs integrated into functional quantum devices. For instance, Gao et al.[254] employed WSe$_2$ SPEs with $g^{(2)}(0)$=0.034 to emulate the BB84 protocol in a QKD setup, achieving click rates of up to 66.95 kHz. Several issues hinder the broader application of TMDC SPEs: most TMDC SPEs require cryogenic temperatures to maintain high *P*; their low *I* due to short dephasing time (~20 ps)[253] constrains their suitability for boson sampling; and the *B* and QY of SPE in TMDC heterostructures are limited due to their electronic band structure and spatial charge separation. Spectral tuning techniques and coupling to photonic structures can help overcome these intrinsic limitations in materials and devices, though much work remains to properly leverage TMDCs in integrated photonic devices for practical quantum applications.

## 8. Spectral Tuning of SPEs

The reproducibility of LD-SPEs is limited by fabrication variability and material inhomogeneity. Spectral tuning can be achieved by strain engineering, the Stark effect, chemical functionalization, or combinations of these approaches. Spectral tuning via strain engineering was demonstrated in SPEs within 2D materials like hBN and TMDCs due to their high Young's modulus (150–400 GPa). Grosso et al.[255] demonstrated that hBN SPEs can achieve spectral tunability of up to 6 meV by strain control using flexible polycarbonate (PC) beams. Xue et al.[256] measured the PL lines of hBN SPEs under varying hydrostatic pressure, demonstrating pressure coefficients of ~15 meV/GPa at 5 K (Fig. 9a & Fig.9b). The SPEs exhibited a flexible bi-directional shift, showing both redshifts and blueshifts in response to pressure applied from different directions. Iff et al.[257] demonstrated a reversible tuning range of up to 18 meV in WSe$_2$ SPEs using piezoelectric actuators (Fig.9c). They observed an energy shift of 5.4 μeV/V by sweeping the electric field applied to the piezoelectric actuator from −20 kV/cm to 20 kV/cm (Fig. 9d).

Spectral tuning via the linear and quadratic Stark effect has been demonstrated in SPEs using hBN, TMDCs, and their heterostructures. The Stark effect is characterized by an energy shift $\Delta\hbar\omega$, which depends on the dipole moment $\vec{\mu}$ and polarizability $\vec{\alpha}$ of the system and is given by $\Delta\hbar\omega = -\Delta\vec{\mu} \cdot \vec{E} - \vec{E} \cdot \left(\frac{\Delta\vec{\alpha}}{2}\right) \cdot \vec{E}$, where $\Delta\vec{\mu}$ is the difference in dipole moments between the excited and ground states and $\Delta\vec{\alpha}$ is the difference in polarizability between these states[258]. The first term represents the linear Stark shift, while the second term reflects the quadratic shift. Chakraborty et al.[259] studied the Stark effect of monolayer $WSe_2$ SPEs in response to vertical electrical field, demonstrating a spectral tunability of up to 21 meV while varying the electric field from -100 MV/m to 400 MV/m. Noh et al.[258] investigated Stark effects of hBN/Graphene SPEs in response to vertical field, demonstrating wavelength shifts as large as 5.4 nm per GV/m. Nikolay et al.[260] studied Stark effect of hBN SPEs in response to vertical field, demonstrating reversible wavelength shifts of $(5.5 \pm 0.3)$ nm around 670nm by applying 20 V. Zhigulin et al.[261] studied Stark effects of blue hBN SPEs in response to lateral fields, demonstrating $|\Delta\vec{\mu}|_{||}$ ~0.1 D and $|\Delta\vec{\alpha}|$~1078 $\text{Å}^3$. Ciarrocchi et al.[262] demonstrated electrical tuning of IXs in $MoSe_2$/$WSe_2$ heterostructures (Fig. 9e), achieving an energy shift from 1.330 eV to 1.468 eV and a $|\Delta\vec{\mu}|_{||}$ ~24 D. Baek et al.[249] observed DC Stark effects in SPEs within $MoSe_2$/$WSe_2$ moiré heterobilayers, achieving a tuning range of ~40 meV and a $|\Delta\vec{\mu}|_{||}$ ~21 D, with $g^{(2)}(0) = 0.28$ at cryogenic temperature.

Spectral tuning via chemical functionalization was demonstrated in SPEs within SWCNTs and TMDCs. He et al.[74] demonstrated tunable wavelengths ranging from 1280nm to 1550nm in SWCNT SPEs at room temperature by varying the chiral index and aryl functionalization. Zhao et al.[33] demonstrated a wavelength shift in GQD SPEs, from approximately 650 nm to 760 nm after functionalization with chlorine atoms. Utama et al.[263] demonstrated a chemomechanical approach to modify the spectra of $WSe_2$ SPEs (Fig. 9f). They applied surface modification to strained monolayer $WSe_2$ using 4-nitrobenzenediazonium (4-NBD) tetrafluoroborate. This process quenched most strain-induced defect emission, resulting in sharp SPEs with $g^{(2)}(0) = 0.01$ after background correction and clean emission spectra and spectral tuning of approximately 20 meV.

## 9. Cavity Coupling for SPE Optimization

Coupling LD-SPEs to optical cavities enhances the **B** and **I** by introducing additional decay pathways and strong light-matter interactions due to electromagnetic confinement. As shown in Fig. 10a, a two-level emitter coupled to an optical cavity mode can dissipate energy into the optical cavity via a Jaynes-Cummings type interaction[264], characterized by a coherent coupling rate ($g$). The cavity then releases the energy into the environment at a rate determined by the cavity linewidth ($\kappa$). The emission enhancement is quantified by the Purcell factor ($F_P$), which is defined as $F_P = \frac{W_{cav}}{W_{free}}$, where $W_{cav}$ and $W_{free}$ are the emission rates in the cavity and free space[103], respectively. The Purcell factor also depends on the quality factor (Q) and the mode volume (V) of the cavity, following the relation $F_P \propto \frac{Q}{V}$. In the weak coupling regime, where g ≪ κ, the Purcell factor is given by $F_P = \frac{4g}{\kappa \gamma_r}$. Cavity quantum electrodynamics (CQED) models for SPEs have been studied and discussed in numerous papers[265,266].

A growing literature has focused on optimizing **P** and **I** in 0D QD SPEs and 1D SWCNT SPEs through integration with optical cavities. For instance, Kim et al.[267] coupled InAs/InP QD SPEs to nanophotonic cavities, demonstrating a visibility of 67% after post-selection. They measured the lifetimes of 32 QDs from 20 cavities, observing a lifetime as short as 400 ps for the fastest-emitting QD, corresponding to a $F_P$ of 4.4 ± 0.5 (Fig. 10b). Tomm et al.[96] coupled InGaAs EQD SPEs to a tunable Fabry-Pérot cavity, featuring a concave top mirror embedded in a silica substrate and a planar bottom mirror in the semiconductor heterostructure. They achieved $g^{(2)}(0) = 0.02$, visibility of 0.98 and an end-to-end efficiency of 0.57. Ding et al.[268] coupled InAs EQD SPEs to a similar microcavity, exhibiting high **P** (0.9795), **I** (0.9856) and efficiency of 0.712 at 4K. Their open Fabry-Pérot cavity consisted of a concave top mirror made from 5.5 pairs of SiO2/Ta2O5 distributed Bragg reflector (DBR) layers, and a bottom mirror featuring a λ-thick quantum dot membrane on a 30-pair AlAs/GaAs DBR, with reflectivities of ~98.6% and ~99.97%, respectively. Farrow et al.[139] coupled CsPbBr3 PQD SPEs to an optical microcavity, realizing **P** of 0.94 and a narrow linewidth of approximately 1 nm. The custom open Fabry-Pérot microcavity features a small mode volume (<1 μm³), high Q-factors (>10⁴), and tunable wavelengths (450–950 nm), consisting of a planar mirror with spin-coated PQDs and a curved mirror, with positioning controlled by piezoelectric stages. Jun et al.[138] coupled CsPbBr3 PQD SPEs to a circular Bragg grating (CBG) cavity made by Si₃N₄ on silicon, realizing ultrafast (<100 ps) single-photon emission and 5.4-fold PL enhancement. . Jeantet et al.[269,270] coupled PFO-wrapped CoMoCat SWCNT SPEs to a fiber microcavity[271] (Fig. 10c), achieving a $F_P$ of up to 60 and a 20-fold PL enhancement at 20K. The cavity features an asymmetric Fabry-Pérot structure with a 50 μm radius of curvature top mirror, mounted on an optical fiber using CO₂ laser ablation, and a back mirror allowing 88% photon transmission. Leveraging the flexibility of the fiber cavity, they achieved a ~10 meV tunable emission energy by adjusting cavity detuning and explored unique exciton-phonon interactions, demonstrating an I of 0.25. Husel et al.[112] coupled SWCNT SPEs to a fiber cavity and operated the system in the regime of incoherent good cavity coupling[272], where photon coherence time is governed by the cavity linewidth. By choosing a cavity with a spectrally narrow linewidth, the SPEs exhibited HOM visibility of 0.65 at room temperature in the telecom band.

2D SPEs can conform closely with optical cavities, leading to a high $F_P$. Tran et al. (2017)[273] demonstrated the deterministic coupling of hBN SPEs to plasmonic nanocavity arrays (Fig. 10d), exhibiting 2.6-fold PL enhancement and $g^{(2)}(0) = 0.29$. The hBN flakes containing pre-characterized SPEs were transferred onto the plasmonic nanoparticle arrays using a wet transfer method ensuring precise placement and strong coupling between the emitters and the plasmonic cavities. Fröch et al.[214] demonstrated the on-chip integration of CVD-grown hBN SPEs with one-dimensional photonic crystal nanobeam cavities fabricated from $Si_3N_4$ (Fig. 10e), achieving a Q of 3300 and a 6-fold PL enhancement. Nonahal et al.[70] coupled blue hBN SPEs to a 1D nanobeam photonic crystal cavity made from exfoliated hBN flakes, achieving a Q over 1000, a $F_P$ of 76 and a 4-fold PL enhancement at room temperature. Cai et al.[274] coupled monolayer $WSe_2$ SPEs with the plasmon mode of a silver nanowire, achieving a lower-bound coupling efficiency of 26% ± 11%. Flatten et al.[275] coupled $WSe_2$ SPEs to open microcavities, achieving a $F_P$ of up to 8 and 5-fold PL enhancement. Their plano-concave microcavity consisted of two silica substrates, each coated with a DBR of 13 $SiO_2/TiO_2$ pairs, providing >99.95% reflectivity at 740 nm. Luo et al. (2018)[235] demonstrated $WSe_2$ SPE - plasmonic cavity systems with a $F_P$ of up to 551. In this setup, the monolayer $WSe_2$ was positioned onto gold pillars, with the hybrid structure flipped onto a smooth, thin layer of gold (Fig. 10f).

## 10. Conclusions and perspectives

LD-SPEs offer high BPI values, various levels of scalability, room-temperature operation, telecom emission, electrically and optically driven emission, and fine spectral tunability. Key hosts for LD-SPEs include QDs, SWCNTs, hBN, and TMDCs, each with unique strengths and challenges based on their synthesis methods and physical properties. CQD SPEs, especially PQD SPEs, demonstrate near-unity QY, room-temperature operation, high $P$ (0.98), and $I$ (0.56), but are limited by low PL intensity due to strong Auger recombination. EQD SPEs show strong BPI values and scalability but are typically restricted to cryogenic temperatures. SWCNT SPEs offer high $P$ (0.69) and $I$ (0.65) at room temperature in the telecom band but are constrained by low QY and strong exciton-exciton annihilation (EEA). hBN SPEs have demonstrated reproducible wavelengths (i.e., 436 nm for blue emitters), high light collection efficiency (98%), and $I$ (0.56) at room temperature, but suffer from low QY and non-telecom emission. TMDC SPEs have achieved high QY (65%), $P$ (0.95), and integration into photonic heterostructures, but they suffer from low $I$ (2%) due to high dephasing rates. Resonant excitation schemes[276] have potential to improve the $I$ of hBN and TMDC SPEs. Spectral tuning using strain engineering, the Stark effect, and chemical functionalization is essential to fine tuning emission wavelengths, and LD-SPEs are especially well suited to manipulation by nanoscale photonic cavities.

SPE arrays that can generate many indistinguishable single photons simultaneously are an essential building block for photonic quantum computing, but none of the LD-SPEs discussed in this manuscript have demonstrated a suitable combination of BPI values to achieve scalable photonic quantum computing. Uniform emitter performance and deterministic emitter placement are key to scaling quantum technologies based on LD-SPEs. However, the literature targeting uniform SPE performance is limited because it is challenging to manipulate the solid-state environment of LD-SPEs at the nanoscale. Blue hBN SPEs exhibit reproducible wavelengths at 436 nm[183], but the $P$ and $I$ vary among emitters. While there is substantial literature exploring deterministic SPE patterning by ion[182]/laser[277] irradiation, these SPEs still exhibit substantial variation from emitter to emitter because of stochastic changes to the defect environment induced by the patterning[185]. Beyond defects and strain fields, IXs confined in the periodic moiré potential across TMDC bilayers present a promising avenue for SPE arrays[278] enabled by the intrinsic spatial reproducibility of the moiré lattice; improved control of clean 2D interfaces will be essential to the realization of moiré quantum photonic platforms.

Many new materials are emerging as candidates for LD-SPEs. 2D metal monochalcogenides (MMCs), such as InSe[279–281], SnS, GaSe[282,283], and GeSe[79,284], exhibit a direct bandgap in both multi-layer and bulk forms[285–287], enabling realizations of SPEs via strain or defect engineering that are compatible integrated photonic platforms.[288] Mixed-dimensional materials[289] combine the features of different dimensionalities, such as the scalability of 0D QDs with the valleytronics of 2D systems[290]. SPEs from such mixed-dimensional heterostructures offer potential for high $P$ and scalability. Atomically thin perovskites[291] exhibit remarkable optoelectronic properties, including high photoluminescence efficiency and tunable emission wavelengths. SPEs can be realized in such materials via techniques used on 2D materials such as electron-beam and femtosecond-laser irradiation. Recently, heterostructures have been shown to generate chiral single-photon emission[292] and improve $P$, offering new opportunities in quantum applications.

Despite substantial research efforts over the past decade and early demonstrations of quantum metrology[6], quantum computing[5], and quantum networking[293], most LD-SPEs have not yet reached the technological maturity of approaches based on QDs or heralded single photons from SPDC or SFWM processes. However, EQD SPEs have begun to reach the necessary thresholds in brightness, purity, and indistinguishability, which has led to a successful demonstration of 5-photon Boson sampling[154]. PQD SPEs and SWCNT SPEs also exhibit strong potential with their high $I$, although their functionality in actual applications has not yet been explicitly demonstrated. Based on recent advanced with these materials, it is reasonable to expect EQD, PQD, and SWCNT SPEs to be integrated into boosted Bell state measurements[294] and repeaters[295] in the near future. The applications of 2D SPEs are more limited due to low $I$, but the $B$ and $P$ achieved to date make them suitable for QKD applications[4], and further improvements in materials processing could unlock 2D SPEs for a broader range of quantum photonic technologies. The substantial progress in engineering LD spin defects over the past several years has unlocked new opportunities for spin-based quantum sensing, with a particular benefit offered by 2D spin defects in hBN that offer a dual purpose of encapsulating environmentally sensitive materials while hosting sensitive probes of local electric and magnetic fields. However, the literature focused on LD spin defects is still relatively new, and much work remains to determine the fundamental limits of these quantum sensors.

Beyond applications relying on emitted single photons, LD-SPEs may replace traditional solid-state SPEs, such as QD and color centers, in cavity-QED systems and enable novel designs that can be used as quantum memories[296–299]. From an engineering perspective, LD-SPEs may be more compatible with integrated photonic cavities, but control over the SPE properties during photonic integration remains a technical challenge in many cases. The future development of LD-SPEs faces challenges in achieving desirable scalability, on-chip integration, and robustness. Ongoing advancements in theoretical research and fabrication techniques offer the potential to optimize LD-SPEs. As innovations in these areas keep enhancing efficiency and functionality, LD-SPEs are expected to unlock new possibilities across a broad range of applications, paving the way for transformative breakthroughs in quantum technologies and beyond.

**FIGURES:**

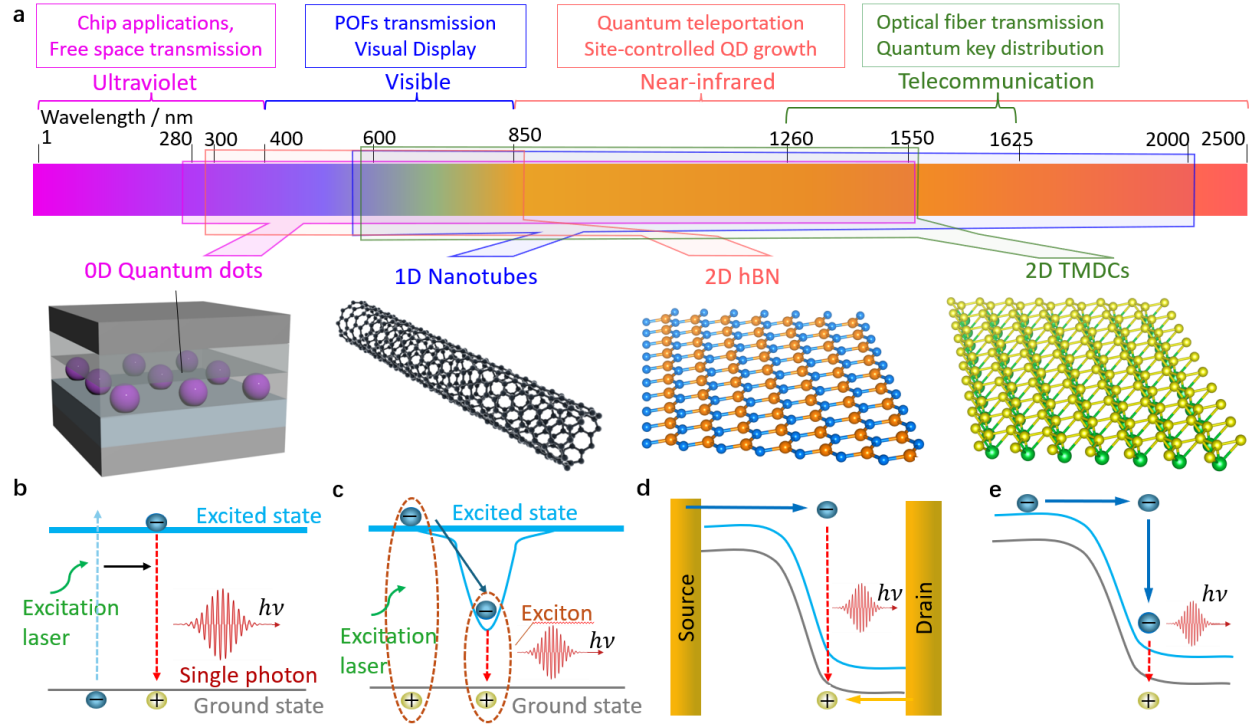

**Figure 1. Wavelength centric overview of low dimensional materials based SPEs**

a) Electromagnetic spectrum showing spectral ranges and applications for SPEs across ultraviolet, visible, near-infrared, and telecommunication wavelengths, from left to right. The top portion highlights applications, while the bottom shows schematic illustrations of SPE materials: quantum dots, nanotubes, hBN, and TMDCs. Colored polygons indicate the spectral ranges covered by each material. b)-e) Different mechanisms for generating single photons. b) Spontaneous decay of excited states, where an excitation laser promotes an electron to the excited state, and single photons are emitted during relaxation. c) Spontaneous decay of localized excitons, where the laser creates excitons, which recombine to emit single photons. d) Ambipolar emission in electroluminescent devices, where electron-hole recombination between the source and drain generates photons. e) Unipolar emission mechanism via impact excitation in electroluminescent devices, where high-energy carriers excite electrons to emit photons during relaxation.

**Table 1: Performance of LD-SPEs in Various Quantum Applications**

| | Purity $g^{(2)}(0)$ | Indistinguishability | Source efficiency |
|---|---|---|---|
| Quantum key distribution[117,195,167] | ~ 0.5 % | --- | ~ 5 % |
| Boson sampling[154,161,162] | ~ 2 % | ~ 94 % | ~ 55 % |
| Quantum teleportation[168,169] | ~ 18 % | ~ 65 % | ~ 15 % |
| Generate cluster state[163-165] | ~ 5 % | ~ 95 % | ~ 18 % |
| Generate GHZ state[114-116] | ~ 2 % | ~ 95 % | ~ 30 % |

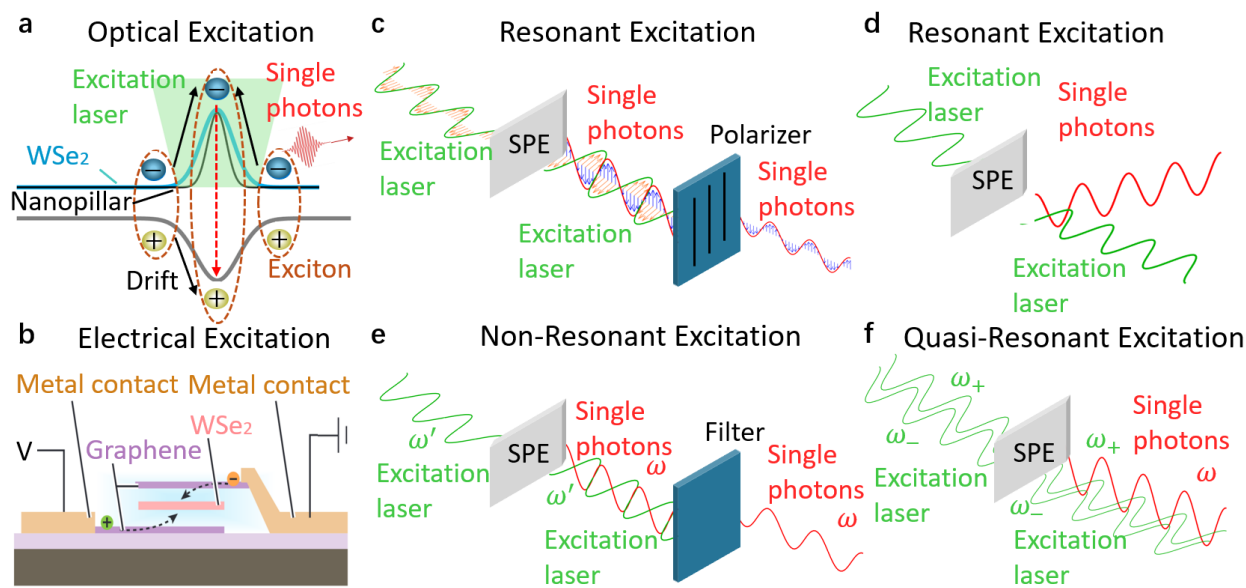

**Figure 2. Different excitation schemes in 2D SPEs.**

a) Schematic of the optical excitation mechanism in a WSe$_2$-based SPEs. An excitation laser excites electrons from the ground state to the excited state, forming excitons in the WSe$_2$ layer positioned on a nanopillar. The excitons drift, and the electron-hole recombination results in the emission of single photons via spontaneous emission.

b) A vertical electrical excitation device comprised of WSe$_2$ sandwiched between few-layer boron nitride and metal contacts. The device structure includes graphene as a conductive layer, with electrical excitation applied through metal contacts to generate excitons within the WSe$_2$ layer.

c) Resonant excitation. The excitation laser and emitted single photons are spectrally filtered based on their polarization.

d) Resonant excitation, spectrally filtering is applied to separate the propagation direction of the excitation laser and the emitted photons.

e) Non-resonant excitation. The excitation laser ($\omega'$) and emitted single photons ($\omega$) have different wavelengths. Spectral filtering is applied to isolate the single photons from the excitation laser based on their distinct wavelengths.

f) A kind of quasi-resonant excitation. Two near-resonant excitation laser pulses ($\omega_+$ and $\omega_-$) are used to predictably manipulate the emitter into its excited state, leading to the emission of single photons ($\omega$).

Adapted with permission from: b, ref.[81] Copyright 2016, American Chemical Society.

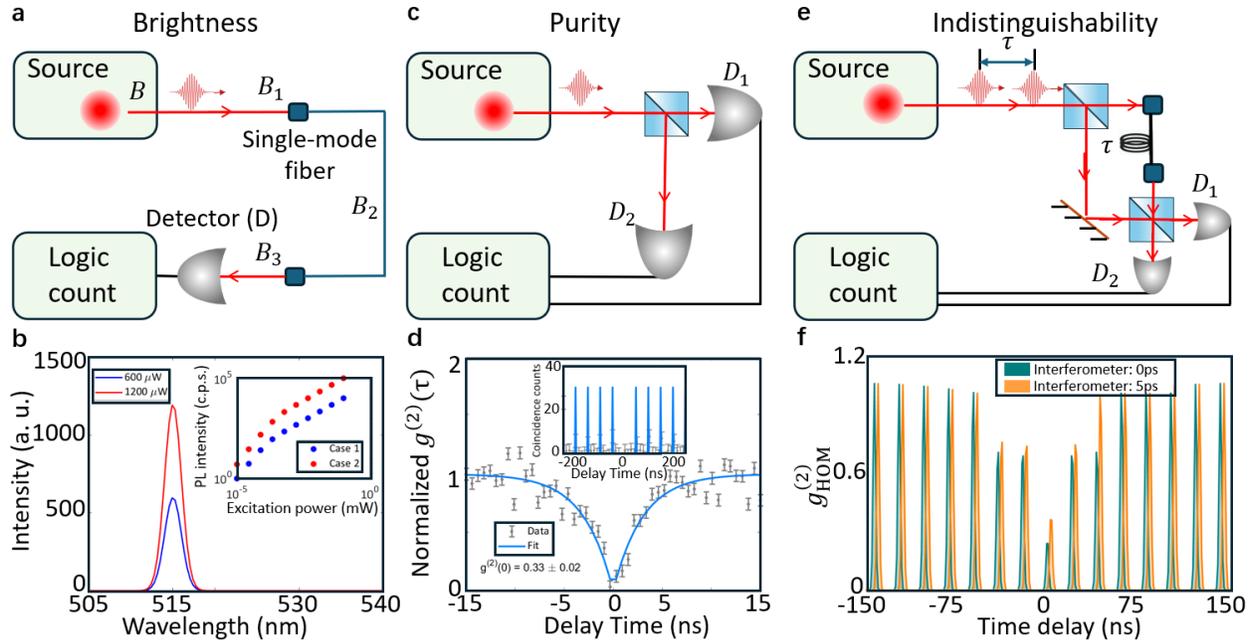

**Figure 3. Criteria of SPEs, purity, brightness and indistinguishability**

a) Experiment set up used to measure brightness **B** of a SPE under pulsed excitation conditions. The emitted photons are collected through a single-mode fiber and detected by a detector (D), with the brightness measured at different points ($B_1$, $B_2$, $B_3$) depending on the collection efficiency. The logic count system is used to record photon detection events.

b) Brightness measurement result showing the integrated PL intensity of a SPE at two excitation powers, 600 μW (blue) and 1200 μW (red), with emission peaks around 515 nm. The inset displays the pump-power-dependent PL intensity of a hBN SPE, highlighting the relationship between excitation power and photon emission for two different cases.

c) HBT experiment setup used to measure purity of an SPE under pulsed excitation conditions. Emitted photons from the source are split by a beam splitter and directed to two detectors, $D_1$ and $D_2$, with the photon coincidences recorded by the logic count system.

d) Single-photon purity measured under continues wave excitation. The data (points) and the fit (blue line) yield $g^{(2)}(0) =0.33\pm0.02$. The inset shows the single-photon purity measured under pulsed excitation, demonstrating clear photon antibunching behavior with well-separated peaks in the coincidence counts.

e) HOM experiment setup used to measure indistinguishability of an SPE under pulsed excitation conditions. Two consecutive photons, separated by a time delay τ, are sent through a beam splitter, and their interference is measured at detectors $D_1$ and $D_2$.

f) Indistinguishability measurement results using the HOM effect, showing the $g^{(2)}_{HOM}(0)$ as a function of time delay for two interferometer settings: 0 ps (dark green) and 5 ps (orange). The reduction in photon correlations at zero-time delay demonstrates two-photon interference.

Reproduced with permission from: a, c, e, ref. [11], John Wiley and Sons; f, ref. [112], Springer Nature

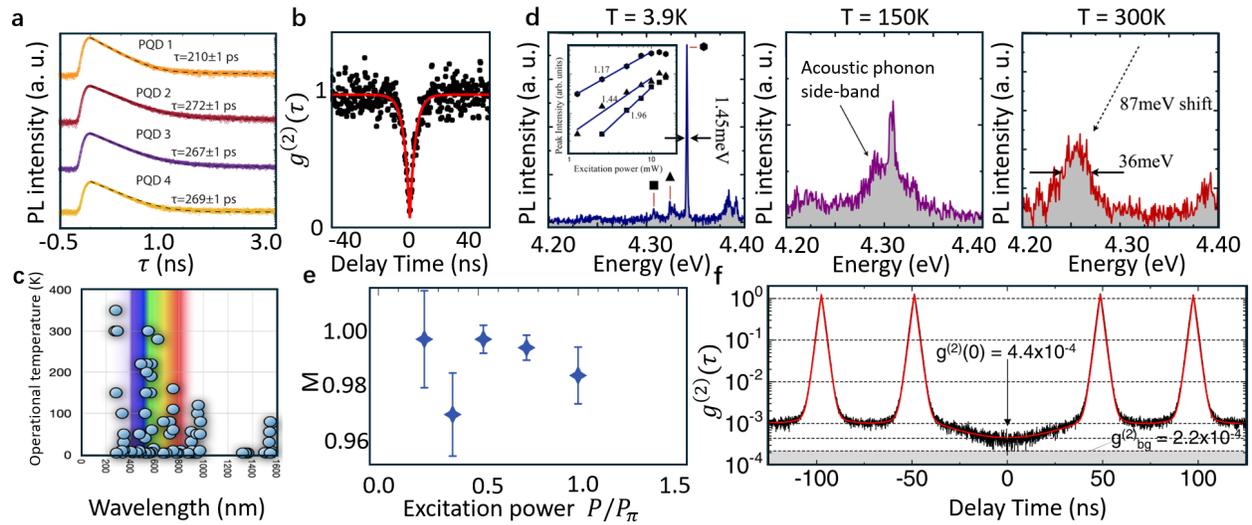

**Figure 4. Single photon emitters based on quantum dots.**

a) The PL decay of a single PQD. The emission exhibits an initial fast decay (~210 to 280 ps), followed by a slower mono-exponential decay spanning two orders of magnitude.

b) Single photon purity of GQD SPEs under non-resonant excitation at room temperature, yielding $g^{(2)}(0)$ equals to $0.05 \pm 0.05$.

c) Summary plot showing emission wavelengths and operational temperatures of various EQD SPEs.

d) PL spectra of EQD SPEs measured at different temperatures: 3.9 K (left), 150 K (middle), and 300 K (right). At 3.9 K, the inset shows a power dependence plot with a fitted slope. At 150 K, the spectrum displays an acoustic phonon sideband. At 300 K, the PL peak exhibits a shift of 87 meV.

e) Indistinguishability M of InGaAs SPEs as a function of excitation power, measured at 4.2 K. The indistinguishability reaches a maximum value of 0.9956. Error bars are based on Poissonian statistics from detected events.

f) The purity of an InAs/InP QD in an optical horn at 8 K under quasi-resonant excitation with a $g^{(2)}(0)$ value of $4.4 \times 10^{-4}$ and a background correction value of $2.2 \times 10^{-4}$.

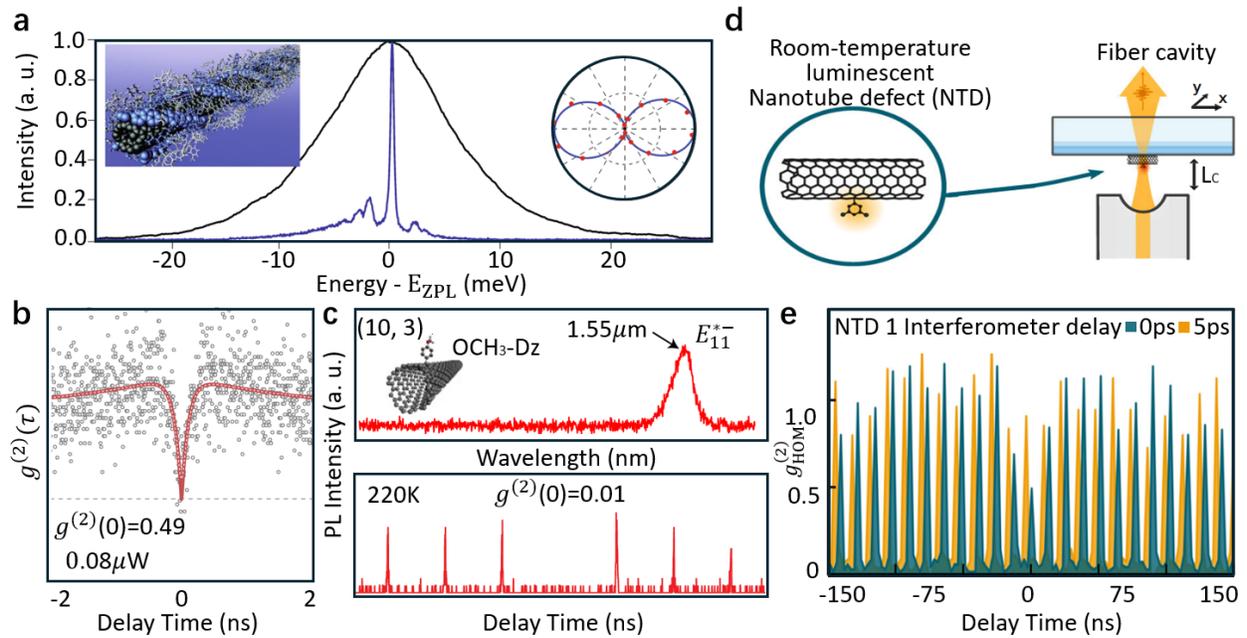

**Figure 5. Single photon emitters based on SWCNTs.**

a) PL spectrum of a single SWCNT at room temperature (black line) and at 10 K (blue line)[173]. Inset: a polymer wrapped nanotube leading to reduced spectral diffusion and blinking[300] (left); PL polarization diagram (right).

b) Purity of electrically excited SWCNT SPEs at 1.6K, yielding $g^{(2)}(0)$ equals to 0.49 at 0.08μW excitation power.

c) PL spectrum and purity for SPEs in (10, 3) SWCNTs functionalized with OCH₃-Dz. The PL spectrum shows an emission peak at 1.55 μm corresponding to the $E_{11}^*$ transition. The $g^{(2)}(0)$ equals to 0.01 measured at 220 K demonstrates high purity of the single-photon emission.

d) Schematic of nanotube defect (NTD) SPEs operating at room temperature and coupled to a tunable fiber cavity. The fiber cavity setup allows precise control of the cavity length ($L_C$) and enhances emission properties of the NTD SPEs.

e) The HOM second-order correlation function of an NTD SPE. HOM autocorrelation function of a NTD, measured in a copolarized interferometer configuration with interferometer delays of 0 ps (dark green) and 5 ps (orange). The zero-interferometer delay corresponds to a delay equal to the separation of one excitation pulse. The visibility is then measured to be v=0.65 ± 0.24 at room temperature.

Adapted with permission from: a, ref. [173], Springer Nature; b, ref. [177], Springer Nature; c, ref. [74], Springer Nature; d, e, ref. [112], Springer Nature.

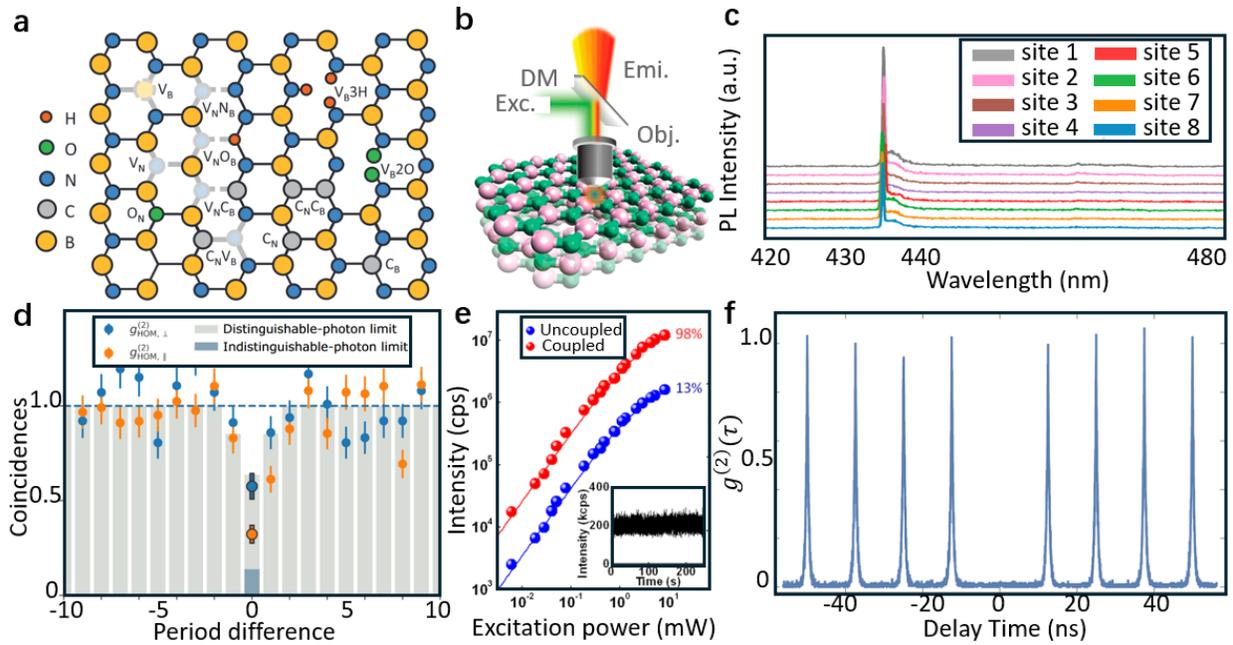

**Figure 6. Structure, defects and performance of hBN.**

a) Schematic of a hBN lattice structure highlighting various types of defects it can host. The lattice is composed of boron (B, yellow) and nitrogen (N, blue) atoms. Common defects include nitrogen vacancies ($V_N$), boron vacancies ($V_B$), oxygen substituting nitrogen or boron ($O_N$, $O_B$), carbon substituting nitrogen or boron ($C_N$, $C_B$), and complex vacancies with multiple atoms missing or substituted (e.g., $V_B3H$, $V_B2O$).

b) Schematic demonstrating the separation of excitation and emission light using a dichroic mirror (DM) during the optical characterization of hBN. The excitation (Exc.) light is directed onto the sample via an objective (Obj.), while the emitted (Emi.) light is reflected by the DM and collected for analysis.

c) Low-temperature spectra of the eight hBN SPEs, labeled 1 through 8. The ZPL for all emitters is reproducible, centered around 436 ± 0.7 nm.

d) Two photon interference between successively emitted photons from the same source with a delay of 12.5 ns, yielding a $V_{HOM}$ of 0.56 ± 0.11.

e) PL intensity versus excitation power for hBN SPEs, comparing uncoupled (blue) and coupled (red) configurations using a metallo-dielectric antenna setup. The coupled system achieves a near-unity photon collection efficiency of 98%, compared to 13% for the uncoupled case. The inset shows the emitter intensity over time, demonstrating excellent temporal stability without blinking.

f) High purity hBN SPEs with $g^{(2)}(0)$ equals to 0.0064 under pulsed excitation.

Adapted with permission from: a, ref. [46], AIP Publishing; b, ref. [67], Copyright 2016, American Chemical Society; c, ref. [183], Springer Nature; d, ref. [195], American Physical Society; e, ref. [93], Copyright 2019, American Chemical Society; f, ref. [106], American Physical Society.

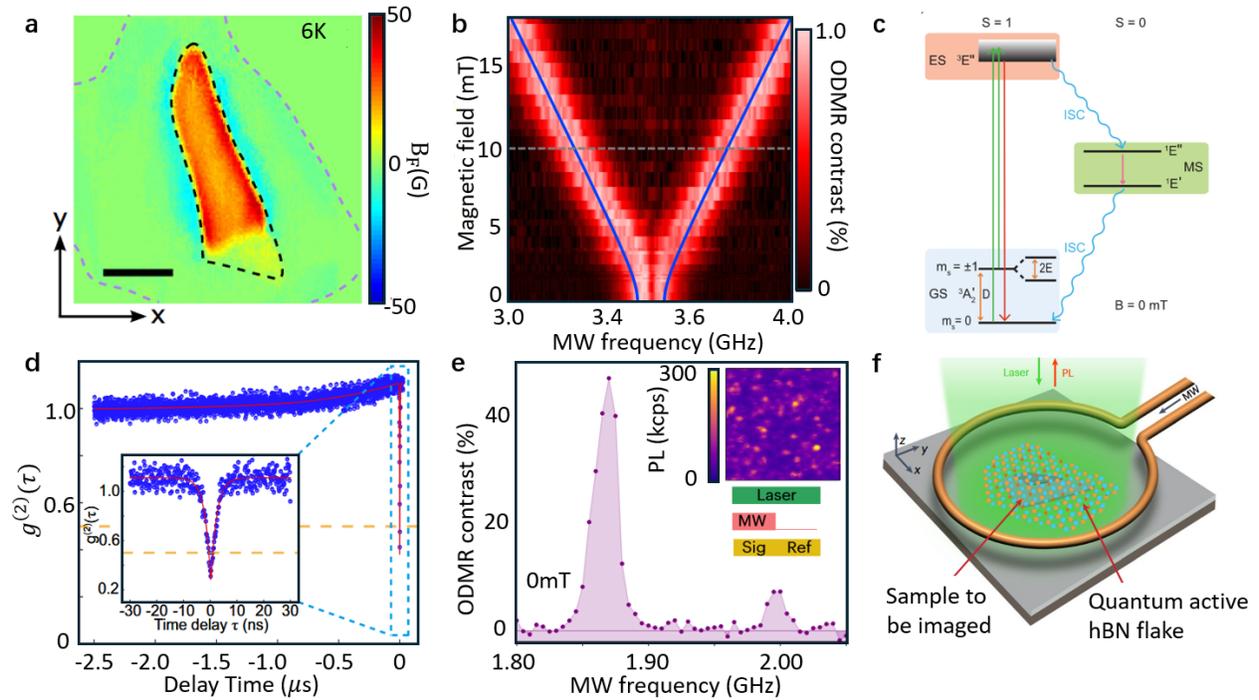

**Figure.7 Quantum spin sensing based on hBN defects**

a) Wide-field imaging of magnetization of an exfoliated $Fe_3GeTe_2$ flake by $V_B^-$ spin defects in hBN.

b) Dependence of ODMR frequencies on the magnetic field. Experimental data (red) and fit (blue line) with parameters D/h = 3.48 GHz, E/h = 50 MHz and g = 2.000.

c) Simplified $V_B^-$ energy-level diagram and the transitions among the ground state ($^3A_2'$), the excited state ($^3E'$), and the metastable state ($^1E'$, $^1E''$).

d) $g^{(2)}(\tau)$ of carbon related defects in hBN, Inset indicates the fitted $g^{(2)}(0)$= 0.25 ± 0.02.

e) The ODMR spectrum of a single defect in hBN measured in the absence of magnetic field. The top-right inset shows a confocal image of the PL intensity of the hBN device under 532 nm laser illumination. The bottom-right inset shows the pulse sequences used in the measurement.

f) A schematic illustration of quantum microscopy with spin defects in hBN. The setup includes a quantum active hBN flake and a sample to be imaged. The laser is used for excitation, and PL is collected for imaging. The microwave (MW) input enables control of the spin defects in the hBN for quantum sensing applications.

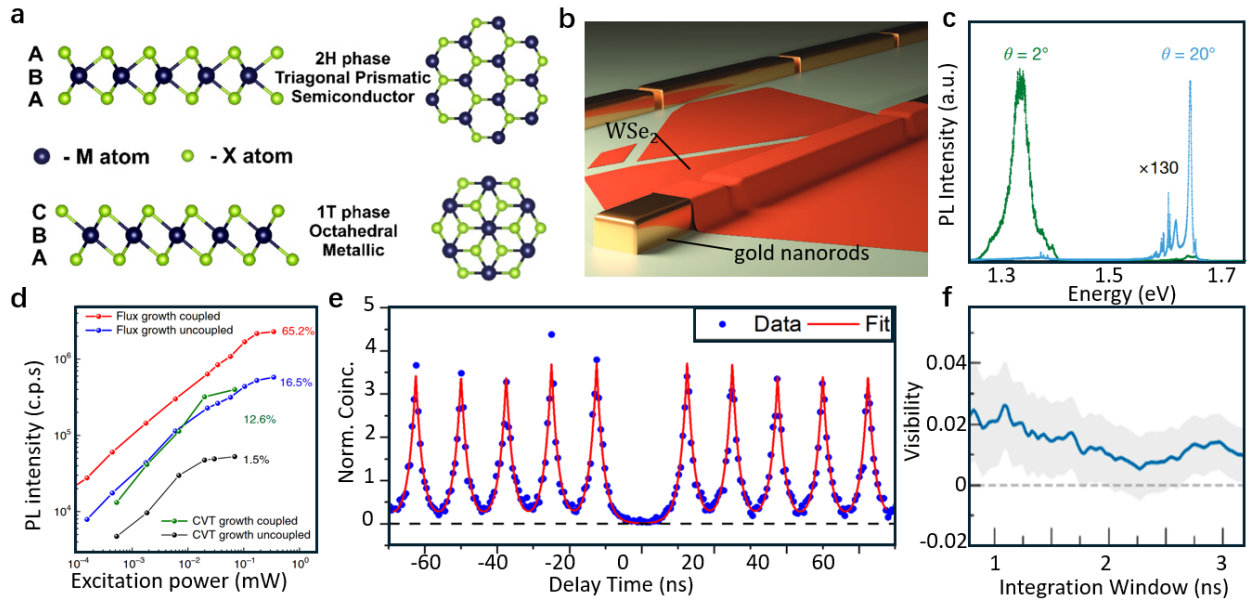

**Figure 8. Structure, strain and performance of layered TMDCs.**

a) Atomic structures of two crystallographic phases of TMDCs. The 2H phase (top) features a trigonal prismatic coordination of metal atoms (M) and chalcogen atoms (X), with an A-B-A stacking sequence. The 1T phase (bottom) is characterized by octahedral coordination and a C-B-A stacking sequence. The side and top views highlight the differences in atomic arrangements between the two phases.

b) Illustration of the WSe$_2$ monolayer transferred over gold nanorods. The strain induced by folds and wrinkles formed during the transfer process, particularly over the gaps between nanorods, leads to the localization of SPEs.

c) PL spectra from MoSe$_2$/WSe$_2$ heterobilayers with twist angles of 2° (green) and 20° (blue; intensity scaled by 130×). The twist angle significantly impacts the PL characteristics, with the 2° sample showing a strong peak near 1.3 eV and the 20° sample exhibiting multiple peaks around 1.6 eV.

d) PL intensity of WSe$_2$ SPEs grown via flux and CVT methods, both before and after coupling to an optical cavity. Flux-grown SPEs show a quantum yield of up to 65.2% after cavity coupling (red), compared to 16.5% without coupling (blue). CVT-grown SPEs achieve a quantum yield of 12.6% with coupling (green) and 1.5% without coupling (black).

e) Second-order autocorrelation function $g^{(2)}(\tau)$ for SPEs in a WSe$_2$ monolayer under pulsed quasi-resonant excitation. The blue data points represent experimental measurements, while the red curve is a fit. The pronounced antibunching at zero delay time, with $g^{(2)}(0)$ equals to 0.036 ± 0.004.

f) HOM interference visibility V$_{HOM}$ as a function of the temporal post-selection window size for SPEs in a WSe$_2$ monolayer coupled to a tunable open optical cavity. The blue line represents the measured visibility, with error bounds in gray. Visibility decreases with increasing integration window size, yielding V$_{HOM}$ equals to 0.02.

Adapted with permission from: a, ref. [223], Licensee MDPI, Basel, Switzerland; b, ref. [301], John Wiley and Sons; c, ref. [248], Springer Nature; d, ref. [235], Springer Nature; e, ref. [252], IOP Publishing; f, ref. [253], American Chemical Society.

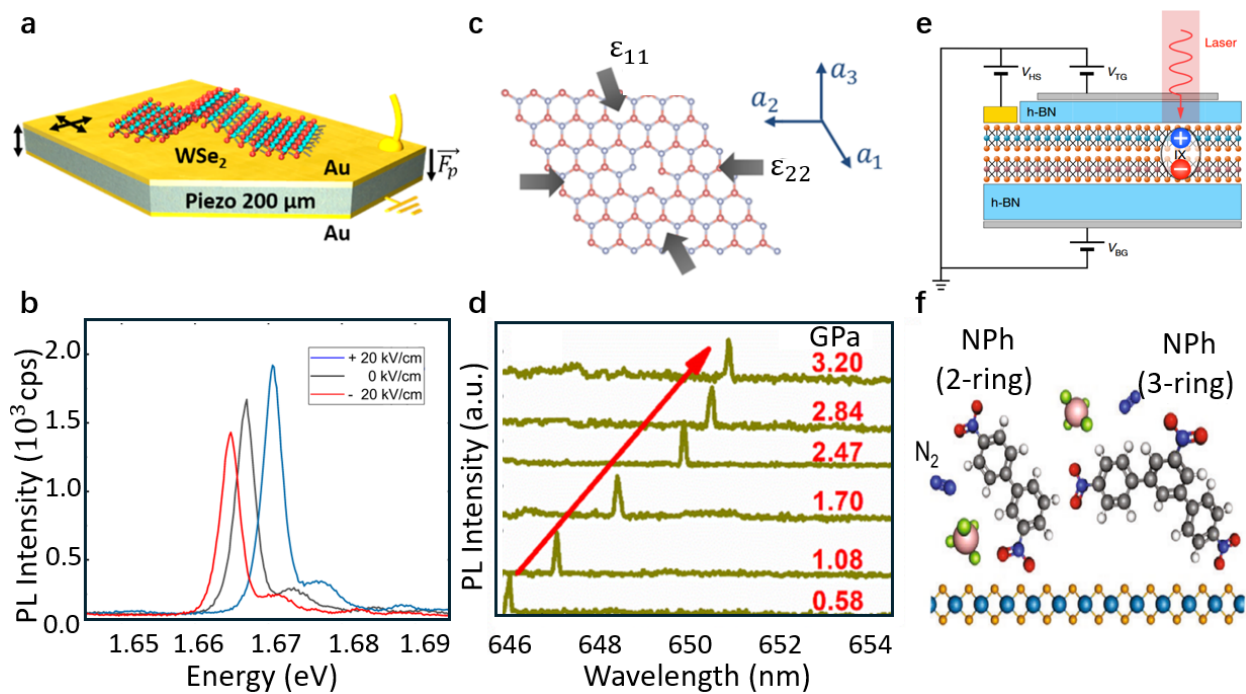

**Figure 9. Spectral tuning of SPEs.**

a) Schematic representation of a 2D hBN flake under strain. The strain components $\varepsilon_{11}$ and $\varepsilon_{22}$ are applied along two principal axes of the lattice, demonstrating the uniaxial or biaxial strain induced in the material. The crystallographic orientation is indicated by the axes $a_1$, $a_2$, and $a_3$.

b) PL spectra of SPEs in the hBN flake under varying pressures, from 0.58 GPa to 3.20 GPa. The PL peak shows a redshift of ~ 5 nm as the applied pressure increases, indicating strain-dependent spectral tuning.

c) Schematic representation of the experimental setup, featuring a WSe$_2$ monolayer placed on a 200 μm piezoelectric substrate. Gold (Au) electrodes are used for electrical contact, and an external voltage is applied across the piezoelectric device to induce strain in the WSe$_2$ layer, enabling spectral tuning.

d) PL spectra of the SPEs in WSe$_2$ monolayer under different applied electric fields: +20 kV/cm (blue), 0 kV/cm (black), and -20 kV/cm (red). The shift in the PL peak energy with varying electric field demonstrates field-dependent control of emission properties.

e) Cross-sectional schematic of a heterostructure device comprising TMDC layers encapsulated by h-BN. The applied gate voltages ($V_{HS}$, $V_{TG}$, and $V_{BG}$) generate an electric field across the TMDC heterostructure, enabling electrical tuning of interlayer excitons (IX) via the DC Stark effect. A laser excites the system, creating interlayer excitons, which are influenced by both the electric field and potential strain (indicated by $F_P$) applied to the device.

f) Illustration of the chemomechanical modification process for SPEs in monolayer WSe$_2$ using aryl diazonium chemistry. Treatment with 4-NBD results in the physisorption of a nitrophenyl (NPh) oligomer layer, consisting of 2-ring and 3-ring structures, onto the WSe$_2$ surface. This functionalization suppresses strain-induced defect emissions, enabling the formation of spectrally isolated SPEs. Nitrogen gas ($N_2$) is released as a by-product of the reaction.

Adapted with permission from: a, b, ref. [256], Copyright 2018, American Chemical Society; c, d, ref. [257], Copyright 2019, American Chemical Society; e, ref. [262], Springer Nature; f, ref. [263], Springer Nature.

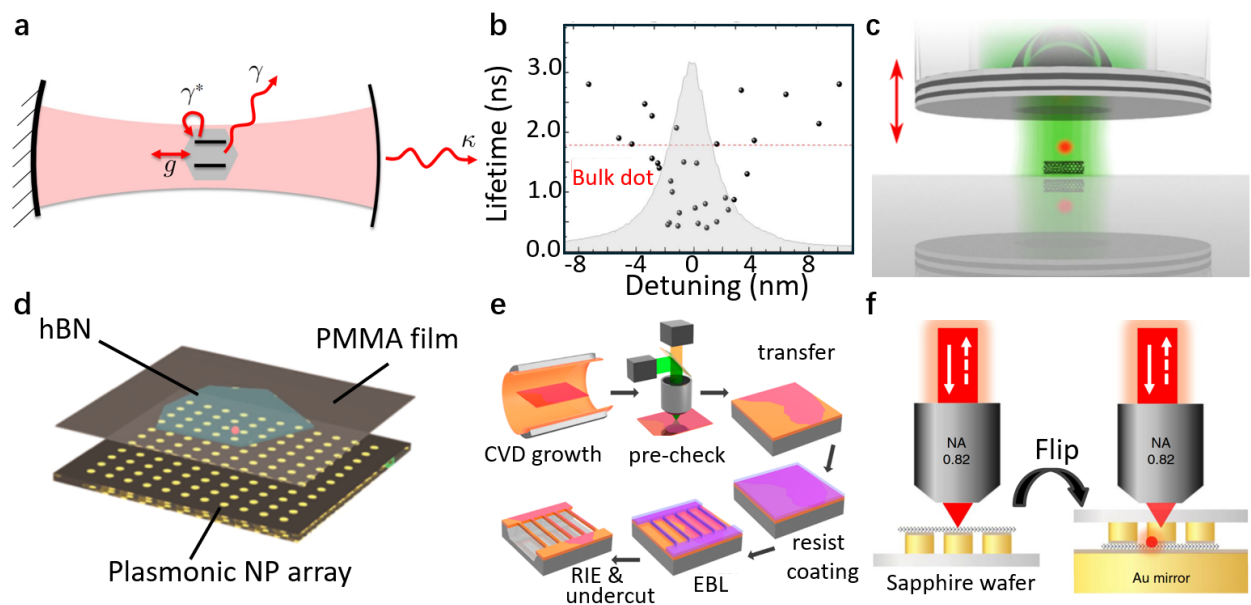

**Figure 10. SPEs couple with plasmonic and photonic cavities.**

a) Schematic of a two-level emitter coupled to an optical cavity mode. The emitter experiences dissipation (γ) and interacts with the cavity mode via coupling strength g. Photons escape the cavity with decay rate κ, while γ∗ represents additional dissipation channels.

b) Lifetime measurements of 32 InAs/InP QDs from 20 different cavities, showing the impact of detuning on the QD lifetime. The shaded region represents the cavity effect, with the red dashed line indicating the lifetime of a bulk dot.

c) Schematic of a SWCNT coupled to a tunable microcavity. The cavity length is adjustable (indicated by the red arrows), allowing control over the optical coupling with the SPEs in the SWCNT.

d) Schematic of a hBN flake placed on a plasmonic nanoparticle array, covered by a poly (methyl methacrylate) (PMMA) film.

e) Fabrication process for integrating CVD-grown hBN with one-dimensional photonic crystal cavities. The steps include CVD growth, transfer, resist coating, electron beam lithography (EBL), and reactive ion etching (RIE) with undercutting.

f) Schematic of the experimental setup used to measure PL intensity and spontaneous emission rate from strain-induced SPEs in a WSe$_2$ monolayer. Left: The emitters are positioned on gold pillars before the formation of the plasmonic nanocavity. Right: After the formation of the plasmonic nanocavity by flipping the material on a planar gold (Au) mirror, leading to enhanced emission properties.

Adapted with permission from: a, ref. [265], Copyright 2015 American Physical Society; b, ref. [267], Copyright 2016 Optical Society of America; c, ref. [270], Copyright 2017 American Chemical Society; d, ref. [273], Copyright 2017, American Chemical Society; e, ref. [214], Copyright 2020, American Chemical Society; f, ref. [235], Springer Nature.

## Reference

1.      J. L. O'Brien, A. Furusawa, and J. Vučković, "Photonic quantum technologies," Nature Photon **3**(12), 687–695 (2009) [doi:10.1038/nphoton.2009.229].

2.      P. Lodahl, S. Mahmoodian, and S. Stobbe, "Interfacing single photons and single quantum dots with photonic nanostructures," Rev. Mod. Phys. **87**(2), 347–400 (2015) [doi:10.1103/RevModPhys.87.347].

3.      J. Lee et al., "Integrated single photon emitters," AVS Quantum Science **2**(3), 031701 (2020) [doi:10.1116/5.0011316].

4.      A. B. D. Shaik and P. Palla, "Optical quantum technologies with hexagonal boron nitride single photon sources," Sci Rep **11**(1), 12285 (2021) [doi:10.1038/s41598-021-90804-4].

5.      M. Gimeno-Segovia et al., "From three-photon greenberger-horne-zeilinger states to ballistic universal quantum computation," Phys. Rev. Lett. **115**(2), 20502 (2015) [doi:10.1103/PhysRevLett.115.020502].

6.      P. Lombardi et al., "Advances in quantum metrology with dielectrically structured single photon sources based on molecules," Adv. Quantum Technol., 2400107 (2024) [doi:10.1002/qute.202400107].

7.      M. Müller et al., "Quantum-dot single-photon sources for entanglement enhanced interferometry," Phys. Rev. Lett. **118**(25), 257402 (2017) [doi:10.1103/PhysRevLett.118.257402].

8.      N. Gisin and R. Thew, "Quantum communication," Nat. Photonics **1**(3), 165–171, Nature Publishing Group (2007) [doi:10.1038/nphoton.2007.22].

9.      C. Kurtsiefer et al., "Stable solid-state source of single photons," Phys. Rev. Lett. **85**(2) (2000).

10.     I. Aharonovich, D. Englund, and M. Toth, "Solid-state single-photon emitters," Nature Photon **10**(10), 631–641 (2016) [doi:10.1038/nphoton.2016.186].

11.     M. Esmann, S. C. Wein, and C. Antón‐Solanas, "Solid‐State Single‐Photon Sources: Recent Advances for Novel Quantum Materials," Adv Funct Materials, 2315936 (2024) [doi:10.1002/adfm.202315936].

12.     X. Ma et al., "Experimental generation of single photons via active multiplexing," Phys. Rev. A **83**(4), 43814 (2011) [doi:10.1103/PhysRevA.83.043814].

13.    J.-W. Pan et al., "Multiphoton entanglement and interferometry," Rev. Mod. Phys. **84**(2), 777–838 (2012) [doi:10.1103/RevModPhys.84.777].

14.    M. Takeoka, R.-B. Jin, and M. Sasaki, "Full analysis of multi-photon pair effects in spontaneous parametric down conversion based photonic quantum information processing," New J. Phys. **17**(4), 43030 (2015) [doi:10.1088/1367-2630/17/4/043030].

15.    A. Chopin et al., "Ultra-efficient generation of time-energy entangled photon pairs in an InGaP photonic crystal cavity," Commun. Phys. **6**(1), 77 (2023) [doi:10.1038/s42005-023-01189-x].

16.    M. J. Holmes et al., "Room-temperature triggered single photon emission from a III-nitride site-controlled nanowire quantum dot," Nano Lett. **14**(2), 982–986 (2014) [doi:10.1021/nl404400d].

17.    Y. Arakawa and M. J. Holmes, "Progress in quantum-dot single photon sources for quantum information technologies: a broad spectrum overview," Appl. Phys. Rev. **7**(2), 21309 (2020) [doi:10.1063/5.0010193].

18.    T. Zhong et al., "Optically addressing single rare-earth ions in a nanophotonic cavity," Phys. Rev. Lett. **121**(18), 183603 (2018) [doi:10.1103/PhysRevLett.121.183603].

19.    J. M. Kindem et al., "Control and single-shot readout of an ion embedded in a nanophotonic cavity," Nature **580**(7802), 201–204 (2020) [doi:10.1038/s41586-020-2160-9].

20.    S. Ourari et al., "Indistinguishable telecom band photons from a single Er ion in the solid state," Nature **620**(7976), 977–981 (2023) [doi:10.1038/s41586-023-06281-4].

21.    A. Dräbenstedt et al., "Low-temperature microscopy and spectroscopy on single defect centers in diamond," Phys. Rev. B **60**(16), 11503–11508 (1999) [doi:10.1103/PhysRevB.60.11503].

22.    N. B. Manson and J. P. Harrison, "Photo-ionization of the nitrogen-vacancy center in diamond," Diamond Relat. Mater. **14**(10), 1705–1710 (2005) [doi:10.1016/j.diamond.2005.06.027].

23.    F. Treussart et al., "Photoluminescence of single colour defects in 50nm diamond nanocrystals," Physica B: Condensed Matter **376–377**, 926–929 (2006) [doi:10.1016/j.physb.2005.12.232].

24.    M. W. Doherty et al., "The nitrogen-vacancy colour centre in diamond," Physics Reports **528**(1), 1–45 (2013) [doi:10.1016/j.physrep.2013.02.001].

25.     R. Schirhagl et al., "Nitrogen-vacancy centers in diamond: nanoscale sensors for physics and biology," Annu. Rev. Phys. Chem. **65**(1), 83–105 (2014) [doi:10.1146/annurev-physchem-040513-103659].

26.     P. Udvarhelyi et al., "Identification of a Telecom Wavelength Single Photon Emitter in Silicon," Phys. Rev. Lett. **127**(19), 196402 (2021) [doi:10.1103/PhysRevLett.127.196402].

27.     M. Hollenbach et al., "Wafer-scale nanofabrication of telecom single-photon emitters in silicon," Nat. Commun. **13**(1), 7683 (2022) [doi:10.1038/s41467-022-35051-5].

28.     N. H. Wan et al., "Efficient Extraction of Light from a Nitrogen-Vacancy Center in a Diamond Parabolic Reflector," Nano Lett. **18**(5), 2787–2793 (2018) [doi:10.1021/acs.nanolett.7b04684].

29.     X. Liu and M. C. Hersam, "2D materials for quantum information science," Nat Rev Mater **4**(10), 669–684 (2019) [doi:10.1038/s41578-019-0136-x].

30.     T. M. Babinec et al., "A diamond nanowire single-photon source," Nat. Nanotechnol. **5**(3), 195–199 (2010) [doi:10.1038/nnano.2010.6].

31.     I. Aharonovich et al., "Diamond-based single-photon emitters," Rep. Prog. Phys. **74**(7), 076501 (2011) [doi:10.1088/0034-4885/74/7/076501].

32.     J. Almutlaq et al., "Closed-loop electron-beam-induced spectroscopy and nanofabrication around individual quantum emitters," Nanophotonics **13**(12), 2251–2258 (2024) [doi:10.1515/nanoph-2023-0877].

33.     S. Zhao et al., "Single photon emission from graphene quantum dots at room temperature," Nat. Commun. **9**(1), 3470 (2018) [doi:10.1038/s41467-018-05888-w].

34.     M. Toth and I. Aharonovich, "Single Photon Sources in Atomically Thin Materials," Annu. Rev. Phys. Chem. **70**(1), 123–142 (2019) [doi:10.1146/annurev-physchem-042018-052628].

35.     S. Gupta et al., "Single-Photon Emission from Two-Dimensional Materials, to a Brighter Future," J. Phys. Chem. Lett. **14**(13), 3274–3284 (2023) [doi:10.1021/acs.jpclett.2c03674].

36.     S. Jesse et al., "Direct atomic fabrication and dopant positioning in Si using electron beams with active real-time image-based feedback," Nanotechnology **29**(25), 255303 (2018) [doi:10.1088/1361-6528/aabb79].

37.     O. Dyck et al., "Atom-by-atom fabrication with electron beams," Nat Rev Mater **4**(7), 497–507 (2019) [doi:10.1038/s41578-019-0118-z].

38.     H. Zhao et al., "Site-controlled telecom-wavelength single-photon emitters in atomically-thin MoTe2," Nat Commun **12**(1), 6753 (2021) [doi:10.1038/s41467-021-27033-w].

39.     S. Häußler et al., "Tunable fiber‐cavity enhanced photon emission from defect centers in hBN," Adv. Opt. Mater. **9**(17), 2002218 (2021) [doi:10.1002/adom.202002218].

40.     X. Xu et al., "Spin and pseudospins in layered transition metal dichalcogenides," Nature Phys **10**(5), 343–350 (2014) [doi:10.1038/nphys2942].

41.     Y. J. Bae et al., "Exciton-coupled coherent magnons in a 2D semiconductor," Nature **609**(7926), 282–286 (2022) [doi:10.1038/s41586-022-05024-1].

42.     N. J. Brennan et al., "Important elements of spin-exciton and magnon-exciton coupling," ACS Phys. Chem. Au **4**(4), 322–327 (2024) [doi:10.1021/acsphyschemau.4c00010].

43.     K. F. Mak et al., "Control of valley polarization in monolayer MoS2 by optical helicity," Nat. Nanotechnol. **7**(8), 494–498 (2012) [doi:10.1038/nnano.2012.96].

44.     J. Yan et al., "Double-pulse generation of indistinguishable single photons with optically controlled polarization," Nano Lett. **22**(4), 1483–1490 (2022) [doi:10.1021/acs.nanolett.1c03543].

45.     S. Ditalia Tchernij et al., "Single-photon emitters in lead-implanted single-crystal diamond," ACS Photonics **5**(12), 4864–4871 (2018) [doi:10.1021/acsphotonics.8b01013].

46.     G. Zhang et al., "Material platforms for defect qubits and single-photon emitters," Applied Physics Reviews **7**(3), 031308 (2020) [doi:10.1063/5.0006075].

47.     J. Wang et al., "Bright room temperature single photon source at telecom range in cubic silicon carbide," Nat Commun **9**(1), 4106 (2018) [doi:10.1038/s41467-018-06605-3].

48.     A. Senichev et al., "Room-temperature single-photon emitters in silicon nitride," Sci. Adv. **7**(50), eabj0627 (2021) [doi:10.1126/sciadv.abj0627].

49.     S. G. Bishop et al., "Room-temperature quantum emitter in aluminum nitride," ACS Photonics **7**(7), 1636–1641 (2020) [doi:10.1021/acsphotonics.0c00528].

50.     Y. Xue et al., "Single-Photon Emission from Point Defects in Aluminum Nitride Films," J. Phys. Chem. Lett. **11**(7), 2689–2694 (2020) [doi:10.1021/acs.jpclett.0c00511].

51.     A. M. Berhane et al., "Bright room‐temperature single‐photon emission from defects in gallium nitride," Adv. Mater. **29**(12), 1605092 (2017) [doi:10.1002/adma.201605092].

52.     Y. Zhou et al., "Room temperature solid-state quantum emitters in the telecom range," SCIENCE ADVANCES (2018).

53.     T. T. Tran et al., "Quantum Emission from Defects in Single-Crystalline Hexagonal Boron Nitride," Phys. Rev. Applied **5**(3), 034005 (2016) [doi:10.1103/PhysRevApplied.5.034005].

54.     T. T. Tran et al., "Quantum emission from hexagonal boron nitride monolayers," Nature Nanotech **11**(1), 37–41 (2016) [doi:10.1038/nnano.2015.242].

55.     A. Srivastava et al., "Optically active quantum dots in monolayer WSe2," Nature Nanotech **10**(6), 491–496 (2015) [doi:10.1038/nnano.2015.60].

56.     M. Koperski et al., "Single photon emitters in exfoliated WSe2 structures," Nature Nanotech **10**(6), 503–506 (2015) [doi:10.1038/nnano.2015.67].

57.     Y.-M. He et al., "Single quantum emitters in monolayer semiconductors," Nature Nanotech **10**(6), 497–502 (2015) [doi:10.1038/nnano.2015.75].

58.     C. Chakraborty et al., "Voltage-controlled quantum light from an atomically thin semiconductor," Nature Nanotech **10**(6), 507–511 (2015) [doi:10.1038/nnano.2015.79].

59.     X. Ma et al., "Room-temperature single-photon generation from solitary dopants of carbon nanotubes," Nat. Nanotechnol. **10**(8), 671–675 (2015) [doi:10.1038/nnano.2015.136].

60.     F. Pyatkov et al., "Cavity-enhanced light emission from electrically driven carbon nanotubes," Nat. Photonics **10**(6), 420–427 (2016) [doi:10.1038/nphoton.2016.70].

61.     M. E. Reimer et al., "Bright single-photon sources in bottom-up tailored nanowires," Nat. Commun. **3**(1), 737 (2012) [doi:10.1038/ncomms1746].

62.     T. B. Hoang, G. M. Akselrod, and M. H. Mikkelsen, "Ultrafast room-temperature single photon emission from quantum dots coupled to plasmonic nanocavities," Nano Lett. **16**(1), 270–275 (2016) [doi:10.1021/acs.nanolett.5b03724].

63.     Z. Yuan et al., "Electrically driven single-photon source," Science **295**(5552), 102–105 (2002) [doi:10.1126/science.1066790].

64.     C. Santori et al., "Indistinguishable photons from a single-photon device," Nature **419**(6907), 594–597 (2002) [doi:10.1038/nature01086].

65.     M. A. Feldman et al., "Evidence of photochromism in a hexagonal boron nitride single-photon emitter," Optica, OPTICA **8**(1), 1–5 (2021) [doi:10.1364/OPTICA.406184].

66.    A. Sajid, M. J. Ford, and J. R. Reimers, "Single-photon emitters in hexagonal boron nitride: a review of progress," Rep. Prog. Phys. **83**(4), 044501 (2020) [doi:10.1088/1361-6633/ab6310].

67.    T. T. Tran et al., "Robust multicolor single photon emission from point defects in hexagonal boron nitride," ACS Nano **10**(8), 7331–7338 (2016) [doi:10.1021/acsnano.6b03602].

68.    L. Gan et al., "Large-scale, high-yield laser fabrication of bright and pure single-photon emitters at room temperature in hexagonal boron nitride," ACS Nano **16**(9), 14254–14261 (2022) [doi:10.1021/acsnano.2c04386].

69.    N. Chejanovsky et al., "Structural Attributes and Photodynamics of Visible Spectrum Quantum Emitters in Hexagonal Boron Nitride," Nano Lett. **16**(11), 7037–7045 (2016) [doi:10.1021/acs.nanolett.6b03268].

70.    M. Nonahal et al., "Deterministic fabrication of a coupled cavity–emitter system in hexagonal boron nitride," Nano Lett. **23**(14), 6645–6650 (2023) [doi:10.1021/acs.nanolett.3c01836].

71.    M. Nguyen et al., "Nanoassembly of quantum emitters in hexagonal boron nitride and gold nanospheres," Nanoscale **10**(5), 2267–2274 (2018) [doi:10.1039/C7NR08249E].

72.    T. Vogl et al., "Fabrication and deterministic transfer of high-quality quantum emitters in hexagonal boron nitride," ACS Photonics **5**(6), 2305–2312 (2018) [doi:10.1021/acsphotonics.8b00127].

73.    Z.-Q. Xu et al., "Single photon emission from plasma treated 2D hexagonal boron nitride," Nanoscale **10**(17), 7957–7965 (2018) [doi:10.1039/C7NR08222C].

74.    X. He et al., "Tunable room-temperature single-photon emission at telecom wavelengths from sp3 defects in carbon nanotubes," Nat. Photonics **11**(9), 577–582 (2017) [doi:10.1038/nphoton.2017.119].

75.    A. Branny et al., "Deterministic strain-induced arrays of quantum emitters in a two-dimensional semiconductor," Nat Commun **8**(1), 15053 (2017) [doi:10.1038/ncomms15053].

76.    C. Chakraborty, N. Vamivakas, and D. Englund, "Advances in quantum light emission from 2D materials," Nanophotonics **8**(11), 2017–2032 (2019) [doi:10.1515/nanoph-2019-0140].

77.     F. Peyskens et al., "Integration of single photon emitters in 2D layered materials with a silicon nitride photonic chip," Nat Commun **10**(1), 4435 (2019) [doi:10.1038/s41467-019-12421-0].

78.     A. Saha et al., "Narrow-band single-photon emission through selective aryl functionalization of zigzag carbon nanotubes," Nat. Chem. **10**(11), 1089–1095 (2018) [doi:10.1038/s41557-018-0126-4].

79.     P. Tonndorf et al., "Single-photon emitters in GaSe," 2D Mater. **4**(2), 021010 (2017) [doi:10.1088/2053-1583/aa525b].

80.     N. R. Jungwirth et al., "Temperature dependence of wavelength selectable zero-phonon emission from single defects in hexagonal boron nitride," Nano Lett. **16**(10), 6052–6057 (2016) [doi:10.1021/acs.nanolett.6b01987].

81.     G. Clark et al., "Single defect light-emitting diode in a van der waals heterostructure," Nano Lett. **16**(6), 3944–3948 (2016) [doi:10.1021/acs.nanolett.6b01580].

82.     W. Luo et al., "Imaging Strain-Localized Single-Photon Emitters in Layered GaSe below the Diffraction Limit," ACS Nano **17**(23), 23455–23465 (2023) [doi:10.1021/acsnano.3c05250].

83.     M. A. Feldman et al., "Colossal photon bunching in quasiparticle-mediated nanodiamond cathodoluminescence," Phys. Rev. B **97**(8), 081404 (2018) [doi:10.1103/PhysRevB.97.081404].

84.     V. Iyer et al., "Photon bunching in cathodoluminescence induced by indirect electron excitation," Nanoscale **15**(22), 9738–9744 (2023) [doi:10.1039/D3NR00376K].

85.     D. Curie et al., "Correlative nanoscale imaging of strained hBN spin defects," ACS Applied Materials & Interfaces **14**(36), 41361–41368 (2022).

86.     S. Roux et al., "Cathodoluminescence monitoring of quantum emitter activation in hexagonal boron nitride," Applied Physics Letters **121**(18), 184002 (2022) [doi:10.1063/5.0126357].

87.     R. Bourrellier et al., "Bright UV single photon emission at point defects in $h$ -BN," Nano Lett. **16**(7), 4317–4321 (2016) [doi:10.1021/acs.nanolett.6b01368].

88.     L. H. G. Tizei and M. Kociak, "Spatially Resolved Quantum Nano-Optics of Single Photons Using an Electron Microscope," Phys. Rev. Lett. **110**(15), 153604 (2013) [doi:10.1103/PhysRevLett.110.153604].

89.     A. Gale et al., "Site-Specific Fabrication of Blue Quantum Emitters in Hexagonal Boron Nitride," ACS Photonics **9**(6), 2170–2177 (2022) [doi:10.1021/acsphotonics.2c00631].

90.     S. Meuret et al., "Photon Bunching in Cathodoluminescence," Phys. Rev. Lett. **114**(19), 197401 (2015) [doi:10.1103/PhysRevLett.114.197401].

91.     M. Solà-Garcia et al., "Photon Statistics of Incoherent Cathodoluminescence with Continuous and Pulsed Electron Beams," ACS Photonics **8**(3), 916–925 (2021) [doi:10.1021/acsphotonics.0c01939].

92.     S. Fiedler et al., "Sub-to-super-Poissonian photon statistics in cathodoluminescence of color center ensembles in isolated diamond crystals," Nanophotonics **12**(12), 2231–2237 (2023) [doi:10.1515/nanoph-2023-0204].

93.     X. Li et al., "Near-unity light collection efficiency from quantum emitters in boron nitride by coupling to metallo-dielectric antennas," ACS Nano **13**(6), 6992–6997 (2019) [doi:10.1021/acsnano.9b01996].

94.     B. D. Mangum et al., "Disentangling the effects of clustering and multi-exciton emission in second-order photon correlation experiments," Opt. Express **21**(6), 7419 (2013) [doi:10.1364/OE.21.007419].

95.     H. Wang et al., "Towards optimal single-photon sources from polarized microcavities," Nat. Photonics **13**(11), 770–775 (2019) [doi:10.1038/s41566-019-0494-3].

96.     N. Tomm et al., "A bright and fast source of coherent single photons," Nat. Nanotechnol. **16**(4), 399–403 (2021) [doi:10.1038/s41565-020-00831-x].

97.     T. T. Tran et al., "Resonant excitation of quantum emitters in hexagonal boron nitride," ACS Photonics **5**(2), 295–300 (2018) [doi:10.1021/acsphotonics.7b00977].

98.     J. M. Raimond, M. Brune, and S. Haroche, "Colloquium: Manipulating quantum entanglement with atoms and photons in a cavity," Rev, Mod, Phys, **73**(3) (2001).

99.     H. Jayakumar et al., "Deterministic photon pairs and coherent optical control of a single quantum dot," Phys. Rev. Lett. **110**(13), 135505 (2013) [doi:10.1103/PhysRevLett.110.135505].

100.    Y. Karli et al., "SUPER scheme in action: experimental demonstration of red-detuned excitation of a quantum emitter," Nano Lett. **22**(16), 6567–6572 (2022) [doi:10.1021/acs.nanolett.2c01783].

101.    F. Sbresny et al., "Stimulated generation of indistinguishable single photons from a quantum ladder system," Phys. Rev. Lett. **128**(9), 93603 (2022) [doi:10.1103/PhysRevLett.128.093603].

102.    Y. Wei et al., "Tailoring solid-state single-photon sources with stimulated emissions," Nat. Nanotechnol. **17**(5), 470–476 (2022) [doi:10.1038/s41565-022-01092-6].

103.    M. Fox, *Quantum optics: an introduction*, Oxford University Press, Oxford ; New York (2006).

104.    C. Jones et al., "Time-dependent Mandel Q parameter analysis for a hexagonal boron nitride single photon source," Opt. Express **31**(6), 10794 (2023) [doi:10.1364/OE.485216].

105.    R. H. Brown and R. Q. Twiss, "Correlation between photons in two coherent beams of light," Nature **177**(4497), 27–29 (1956) [doi:10.1038/177027a0].

106.    T. Vogl et al., "Sensitive single-photon test of extended quantum theory with two-dimensional hexagonal boron nitride," Phys. Rev. Res. **3**(1), 13296 (2021) [doi:10.1103/PhysRevResearch.3.013296].

107.    C. K. Hong, Z. Y. Ou, and L. Mandel, "Measurement of subpicosecond time intervals between two photons by interference," Phys. Rev. Lett. **59**(18), 2044–2046 (1987) [doi:10.1103/PhysRevLett.59.2044].

108.    P. J. Mosley et al., "Heralded generation of ultrafast single photons in pure quantum states," Phys. Rev. Lett. **100**(13), 133601 (2008) [doi:10.1103/PhysRevLett.100.133601].

109.    A. V. Kuhlmann et al., "Charge noise and spin noise in a semiconductor quantum device," Nat. Phys. **9**(9), 570–575 (2013) [doi:10.1038/nphys2688].

110.    P. Borri et al., "Ultralong Dephasing Time in InGaAs Quantum Dots," Phys. Rev. Lett. **87**(15), 157401 (2001) [doi:10.1103/PhysRevLett.87.157401].

111.    H. Utzat et al., "Coherent single-photon emission from colloidal lead halide perovskite quantum dots," Science **363**(6431), 1068–1072 (2019) [doi:10.1126/science.aau7392].

112.    L. Husel et al., "Cavity-enhanced photon indistinguishability at room temperature and telecom wavelengths," Nat. Commun. **15**(1), 3989 (2024) [doi:10.1038/s41467-024-48119-1].

113.    C. H. Bennett and G. Brassard, "Quantum cryptography: public key distribution and coin tossing," Theor. Comput. Sci. **560**, 7–11 (2014) [doi:10.1016/j.tcs.2014.05.025].

114.    M. H. Appel et al., "Entangling a Hole Spin with a Time-Bin Photon: A Waveguide Approach for Quantum Dot Sources of Multiphoton Entanglement," Phys. Rev. Lett. **128**(23), 233602 (2022) [doi:10.1103/PhysRevLett.128.233602].

115.    K. Tiurev et al., "High-fidelity multiphoton-entangled cluster state with solid-state quantum emitters in photonic nanostructures," Phys. Rev. A **105**(3), L030601 (2022) [doi:10.1103/PhysRevA.105.L030601].

116.    H. Cao et al., "Photonic Source of Heralded Greenberger-Horne-Zeilinger States," Phys. Rev. Lett. **132**(13), 130604 (2024) [doi:10.1103/PhysRevLett.132.130604].

117.    K. Takemoto et al., "Quantum key distribution over 120 km using ultrahigh purity single-photon source and superconducting single-photon detectors," Sci. Rep. **5**(1), 14383 (2015) [doi:10.1038/srep14383].

118.    Y. Lin, Y. Ye, and W. Fang, "Electrically driven single-photon sources," J. Semicond. **40**(7), 071904 (2019) [doi:10.1088/1674-4926/40/7/071904].

119.    M. J. Holmes et al., "Single Photons from a Hot Solid-State Emitter at 350 K," ACS Photonics **3**(4), 543–546 (2016) [doi:10.1021/acsphotonics.6b00112].

120.    M. Munsch et al., "Dielectric GaAs antenna ensuring an efficient broadband coupling between an InAs quantum dot and a gaussian optical beam," Phys. Rev. Lett. **110**(17), 177402 (2013) [doi:10.1103/PhysRevLett.110.177402].

121.    M. Liu et al., "Colloidal quantum dot electronics," Nat. Electron. **4**(8), 548–558 (2021) [doi:10.1038/s41928-021-00632-7].

122.    Y. Yan et al., "Recent Advances on Graphene Quantum Dots: From Chemistry and Physics to Applications," Advanced Materials **31**(21), 1808283 (2019) [doi:10.1002/adma.201808283].

123.    C. B. Murray, D. J. Norris, and M. G. Bawendi, "Synthesis and characterization of nearly monodisperse CdE (E = sulfur, selenium, tellurium) semiconductor nanocrystallites," J. Am. Chem. Soc. **115**(19), 8706–8715 (1993) [doi:10.1021/ja00072a025].

124.    V. Chandrasekaran et al., "Nearly blinking-free, high-purity single-photon emission by colloidal InP/ZnSe quantum dots," Nano Lett. **17**(10), 6104–6109 (2017) [doi:10.1021/acs.nanolett.7b02634].

125.    A. Jain et al., "Atomistic design of CdSe/CdS core–shell quantum dots with suppressed auger recombination," Nano Lett. **16**(10), 6491–6496 (2016) [doi:10.1021/acs.nanolett.6b03059].

126.     S. Krishnamurthy et al., "PbS/CdS Quantum Dot Room-Temperature Single-Emitter Spectroscopy Reaches the Telecom O and S Bands via an Engineered Stability," ACS Nano **15**(1), 575–587 (2021) [doi:10.1021/acsnano.0c05907].

127.     J. Zou et al., "Perovskite quantum dots: synthesis, applications, prospects, and challenges," J. Appl. Phys. **132**(22), 220901 (2022) [doi:10.1063/5.0126496].

128.     L. Protesescu et al., "Nanocrystals of Cesium Lead Halide Perovskites (CsPbX $_3$ , X = Cl, Br, and I): Novel Optoelectronic Materials Showing Bright Emission with Wide Color Gamut," Nano Lett. **15**(6), 3692–3696 (2015) [doi:10.1021/nl5048779].

129.     J. Pan et al., "Highly efficient perovskite‐quantum‐dot light‐emitting diodes by surface engineering," Adv. Mater. **28**(39), 8718–8725 (2016) [doi:10.1002/adma.201600784].

130.     F. Zhang et al., "Colloidal synthesis of air-stable $CH_3NH_3PbI_3$ quantum dots by gaining chemical insight into the solvent effects," Chem. Mater. **29**(8), 3793–3799 (2017) [doi:10.1021/acs.chemmater.7b01100].

131.     I. Levchuk et al., "Brightly luminescent and color-tunable formamidinium lead halide perovskite $FAPbX_3$ (X = Cl, Br, I) colloidal nanocrystals," Nano Lett. **17**(5), 2765–2770 (2017) [doi:10.1021/acs.nanolett.6b04781].

132.     Y. Tong et al., "Highly Luminescent Cesium Lead Halide Perovskite Nanocrystals with Tunable Composition and Thickness by Ultrasonication," Angew Chem Int Ed **55**(44), 13887–13892 (2016) [doi:10.1002/anie.201605909].

133.     L. Rao et al., "Polar-solvent-free synthesis of highly photoluminescent and stable $CsPbBr_3$ nanocrystals with controlled shape and size by ultrasonication," Chem. Mater. **31**(2), 365–375 (2019) [doi:10.1021/acs.chemmater.8b03298].

134.     F. Liu et al., "Highly Luminescent Phase-Stable $CsPbI_3$ Perovskite Quantum Dots Achieving Near 100% Absolute Photoluminescence Quantum Yield," ACS Nano **11**(10), 10373–10383 (2017) [doi:10.1021/acsnano.7b05442].

135.     X. Tang et al., "Single Halide Perovskite/Semiconductor Core/Shell Quantum Dots with Ultrastability and Nonblinking Properties," Advanced Science **6**(18), 1900412 (2019) [doi:10.1002/advs.201900412].

136.     C. Zhu et al., "Room-Temperature, Highly Pure Single-Photon Sources from All-Inorganic Lead Halide Perovskite Quantum Dots," Nano Lett. **22**(9), 3751–3760 (2022) [doi:10.1021/acs.nanolett.2c00756].

137.    A. E. K. Kaplan et al., "Hong–Ou–Mandel interference in colloidal CsPbBr3 perovskite nanocrystals," Nat. Photon. **17**(9), 775–780 (2023) [doi:10.1038/s41566-023-01225-w].

138.    S. Jun et al., "Ultrafast and Bright Quantum Emitters from the Cavity-Coupled Single Perovskite Nanocrystals," ACS Nano **18**(2), 1396–1403 (2024) [doi:10.1021/acsnano.3c06760].

139.    T. Farrow et al., "Ultranarrow Line Width Room-Temperature Single-Photon Source from Perovskite Quantum Dot Embedded in Optical Microcavity," Nano Lett. **23**(23), 10667–10673 (2023) [doi:10.1021/acs.nanolett.3c02058].

140.    J. Wu, W. Pisula, and K. Müllen, "Graphenes as potential material for electronics," Chem. Rev. **107**(3), 718–747 (2007) [doi:10.1021/cr068010r].

141.    H. Arab, S. MohammadNejad, and P. MohammadNejad, "Se-doped NH2-functionalized graphene quantum dot for single-photon emission at free-space quantum communication wavelength," Quantum Inf Process **20**(5), 184 (2021) [doi:10.1007/s11128-021-03122-z].

142.    A. Jasik et al., "The influence of the growth rate and V/III ratio on the crystal quality of InGaAs/GaAs QW structures grown by MBE and MOCVD methods," J. Cryst. Growth **311**(19), 4423–4432 (2009) [doi:10.1016/j.jcrysgro.2009.07.032].

143.    J. Gérard et al., "Enhanced spontaneous emission by quantum boxes in a monolithic optical microcavity," Phys. Rev. Lett. **81**(5), 1110–1113 (1998) [doi:10.1103/PhysRevLett.81.1110].

144.    K. Hennessy et al., "Quantum nature of a strongly coupled single quantum dot–cavity system," Nature **445**(7130), 896–899 (2007) [doi:10.1038/nature05586].

145.    M. J. Holmes, M. Arita, and Y. Arakawa, "III-nitride quantum dots as single photon emitters," Semicond. Sci. Technol. **34**(3), 33001 (2019) [doi:10.1088/1361-6641/ab02c8].

146.    S. Bounouar et al., "Ultrafast room temperature single-photon source from nanowire-quantum dots," Nano Lett. **12**(6), 2977–2981 (2012) [doi:10.1021/nl300733f].

147.    O. Fedorych et al., "Room temperature single photon emission from an epitaxially grown quantum dot," Appl. Phys. Lett. **100**(6), 61114 (2012) [doi:10.1063/1.3683498].

148.    W. Quitsch et al., "Electrically driven single photon emission from a CdSe/ZnSSe/MgS semiconductor quantum dot," Phys. Status Solidi C **11**(7–8), 1256–1259 (2014) [doi:10.1002/pssc.201300627].

149.    S. Deshpande et al., "Electrically driven polarized single-photon emission from an InGaN quantum dot in a GaN nanowire," Nat. Commun. **4**(1), 1675 (2013) [doi:10.1038/ncomms2691].

150.    S. Deshpande et al., "Electrically pumped single-photon emission at room temperature from a single InGaN/GaN quantum dot," Appl. Phys. Lett. **105**(14), 141109 (2014) [doi:10.1063/1.4897640].

151.    J.-H. Cho et al., "Strongly coherent single-photon emission from site-controlled InGaN quantum dots embedded in GaN nanopyramids," ACS Photonics **5**(2), 439–444 (2018) [doi:10.1021/acsphotonics.7b00922].

152.    Y.-M. He et al., "On-demand semiconductor single-photon source with near-unity indistinguishability," Nat. Nanotechnol. **8**(3), 213–217 (2013) [doi:10.1038/nnano.2012.262].

153.    N. Somaschi et al., "Near-optimal single-photon sources in the solid state," Nat. Photonics **10**(5), 340–345 (2016) [doi:10.1038/nphoton.2016.23].

154.    H. Wang et al., "High-efficiency multiphoton boson sampling," Nature Photon **11**(6), 361–365 (2017) [doi:10.1038/nphoton.2017.63].

155.    L. Schweickert et al., "On-demand generation of background-free single photons from a solid-state source," Appl. Phys. Lett. **112**(9), 93106 (2018) [doi:10.1063/1.5020038].

156.    M. Paul et al., "Single-photon emission at 1.55 $\mu$ m from MOVPE-grown InAs quantum dots on InGaAs/GaAs metamorphic buffers," Applied Physics Letters **111**(3), 033102 (2017) [doi:10.1063/1.4993935].

157.    K. Takemoto et al., "An optical horn structure for single-photon source using quantum dots at telecommunication wavelength," J. Appl. Phys. **101**(8), 81720 (2007) [doi:10.1063/1.2723177].

158.    T. Miyazawa et al., "Single-photon emission at 1.5 $\mu$ m from an InAs/InP quantum dot with highly suppressed multi-photon emission probabilities," Appl. Phys. Lett. **109**(13), 132106 (2016) [doi:10.1063/1.4961888].

159.    T. Müller et al., "A quantum light-emitting diode for the standard telecom window around 1,550 nm," Nat Commun **9**(1), 862 (2018) [doi:10.1038/s41467-018-03251-7].

160.    C. Nawrath et al., "Coherence and indistinguishability of highly pure single photons from non-resonantly and resonantly excited telecom C-band quantum dots," Appl. Phys. Lett. **115**(2), 23103 (2019) [doi:10.1063/1.5095196].

161.    R. García-Patrón, J. J. Renema, and V. Shchesnovich, "Simulating boson sampling in lossy architectures," Quantum **3**, 169 (2019) [doi:10.22331/q-2019-08-05-169].

162.    N. Maring et al., "A versatile single-photon-based quantum computing platform," Nat. Photon. **18**(6), 603–609 (2024) [doi:10.1038/s41566-024-01403-4].

163.    D. Istrati et al., "Sequential generation of linear cluster states from a single photon emitter," Nat Commun **11**(1), 5501 (2020) [doi:10.1038/s41467-020-19341-4].

164.    D. Cogan et al., "Deterministic generation of indistinguishable photons in a cluster state," Nat. Photon. (2023) [doi:10.1038/s41566-022-01152-2].

165.    N. Coste et al., "High-rate entanglement between a semiconductor spin and indistinguishable photons," Nat. Photon. **17**(7), 582–587 (2023) [doi:10.1038/s41566-023-01186-0].

166.    H.-K. Lo, M. Curty, and K. Tamaki, "Secure quantum key distribution," Nat. Photonics **8**(8), 595–604 (2014) [doi:10.1038/nphoton.2014.149].

167.    M. A. Pooley et al., "Controlled-NOT gate operating with single photons," Appl. Phys. Lett. **100**(21), 211103 (2012) [doi:10.1063/1.4719077].

168.    M. Reindl et al., "All-photonic quantum teleportation using on-demand solid-state quantum emitters," Sci. Adv. **4**(12), eaau1255 (2018) [doi:10.1126/sciadv.aau1255].

169.    M. Anderson et al., "Quantum teleportation using highly coherent emission from telecom C-band quantum dots," npj Quantum Inf. **6**(1), 14 (2020) [doi:10.1038/s41534-020-0249-5].

170.    D. Golberg et al., "Boron Nitride Nanotubes and Nanosheets," ACS Nano **4**(6), 2979–2993 (2010) [doi:10.1021/nn1006495].

171.    J. Ahn et al., "Stable emission and fast optical modulation of quantum emitters in boron nitride nanotubes," Opt. Lett. **43**(15), 3778 (2018) [doi:10.1364/OL.43.003778].

172.    X. Gao et al., "Nanotube spin defects for omnidirectional magnetic field sensing," Nat. Commun. **15**(1), 7697 (2024) [doi:10.1038/s41467-024-51941-2].

173.    X. He et al., "Carbon nanotubes as emerging quantum-light sources," Nat. Mater. **17**(8), 663–670 (2018) [doi:10.1038/s41563-018-0109-2].

174.    C. Raynaud et al., "Superlocalization of excitons in carbon nanotubes at cryogenic temperature," Nano Lett. **19**(10), 7210–7216 (2019) [doi:10.1021/acs.nanolett.9b02816].

175.    Z. Li et al., "Quantum emission assisted by energy landscape modification in pentacene-decorated carbon nanotubes," ACS Photonics **8**(8), 2367–2374 (2021) [doi:10.1021/acsphotonics.1c00539].

176.    A. Högele et al., "Photon antibunching in the photoluminescence spectra of a single carbon nanotube," Phys. Rev. Lett. **100**(21), 217401 (2008) [doi:10.1103/PhysRevLett.100.217401].

177.    S. Khasminskaya et al., "Fully integrated quantum photonic circuit with an electrically driven light source," Nat. Photonics **10**(11), 727–732 (2016) [doi:10.1038/nphoton.2016.178].

178.    Y. Zheng et al., "Quantum light emission from coupled defect states in DNA-functionalized carbon nanotubes," ACS Nano **15**(6), 10406–10414 (2021) [doi:10.1021/acsnano.1c02709].

179.    F. Hayee et al., "Revealing multiple classes of stable quantum emitters in hexagonal boron nitride with correlated optical and electron microscopy," Nat. Mater. **19**(5), 534–539 (2020) [doi:10.1038/s41563-020-0616-9].

180.    T. T. Tran et al., "Room‐temperature single‐photon emission from oxidized tungsten disulfide multilayers," Adv. Opt. Mater. **5**(5), 1600939 (2017) [doi:10.1002/adom.201600939].

181.    H. Ngoc My Duong et al., "Effects of high-energy electron irradiation on quantum emitters in hexagonal boron nitride," ACS Appl. Mater. Interfaces **10**(29), 24886–24891 (2018) [doi:10.1021/acsami.8b07506].

182.    J. Ziegler et al., "Deterministic Quantum Emitter Formation in Hexagonal Boron Nitride via Controlled Edge Creation," Nano Lett. **19**(3), 2121–2127 (2019) [doi:10.1021/acs.nanolett.9b00357].

183.    C. Fournier et al., "Position-controlled quantum emitters with reproducible emission wavelength in hexagonal boron nitride," Nat Commun **12**(1), 3779 (2021) [doi:10.1038/s41467-021-24019-6].

184.    N. V. Proscia et al., "Near-deterministic activation of room-temperature quantum emitters in hexagonal boron nitride," Optica **5**(9), 1128 (2018) [doi:10.1364/OPTICA.5.001128].

185.    C. Li et al., "Scalable and deterministic fabrication of quantum emitter arrays from hexagonal boron nitride," Nano Lett. **21**(8), 3626–3632 (2021) [doi:10.1021/acs.nanolett.1c00685].

186.    S. Castelletto et al., "Color centers enabled by direct femto-second laser writing in wide bandgap semiconductors," Nanomaterials **11**(1), 72 (2020) [doi:10.3390/nano11010072].

187.    S. A. Tawfik et al., "First-principles investigation of quantum emission from hBN defects," Nanoscale **9**(36), 13575–13582 (2017) [doi:10.1039/C7NR04270A].

188.    P. Auburger and A. Gali, "Towards *ab initio* identification of paramagnetic substitutional carbon defects in hexagonal boron nitride acting as quantum bits," Phys. Rev. B **104**(7), 75410 (2021) [doi:10.1103/PhysRevB.104.075410].

189.    Q. Tan et al., "Donor–Acceptor Pair Quantum Emitters in Hexagonal Boron Nitride," Nano Lett. **22**(3), 1331–1337 (2022) [doi:10.1021/acs.nanolett.1c04647].

190.    Á. Ganyecz et al., "First-principles theory of the nitrogen interstitial in hBN: a plausible model for the blue emitter," Nanoscale **16**(8), 4125–4139 (2024) [doi:10.1039/D3NR05811E].

191.    C. Jara et al., "First-Principles Identification of Single Photon Emitters Based on Carbon Clusters in Hexagonal Boron Nitride," J. Phys. Chem. A **125**(6), 1325–1335 (2021) [doi:10.1021/acs.jpca.0c07339].

192.    O. Golami et al., "A b i n i t i o and group theoretical study of properties of a carbon trimer defect in hexagonal boron nitride," Phys. Rev. B **105**(18), 184101 (2022) [doi:10.1103/PhysRevB.105.184101].

193.    S. Li et al., "Ultraviolet Quantum Emitters in Hexagonal Boron Nitride from Carbon Clusters," J. Phys. Chem. Lett. **13**(14), 3150–3157 (2022) [doi:10.1021/acs.jpclett.2c00665].

194.    J. Horder et al., "Coherence Properties of Electron-Beam-Activated Emitters in Hexagonal Boron Nitride Under Resonant Excitation," Phys. Rev. Applied **18**(6), 064021 (2022) [doi:10.1103/PhysRevApplied.18.064021].

195.    C. Fournier et al., "Two-photon interference from a quantum emitter in hexagonal boron nitride," Phys. Rev. Appl. **19**(4), L041003 (2023) [doi:10.1103/PhysRevApplied.19.L041003].

196.    T. Vogl et al., "Compact cavity-enhanced single-photon generation with hexagonal boron nitride," ACS Photonics **6**(8), 1955–1962 (2019) [doi:10.1021/acsphotonics.9b00314].

197.    S. J. U. White et al., "Quantum random number generation using a hexagonal boron nitride single photon emitter," J. Opt. **23**(1), 1LT01 (2021) [doi:10.1088/2040-8986/abccff].

198.    Ç. Samaner et al., "Free‐space quantum key distribution with single photons from defects in hexagonal boron nitride," Adv. Quantum Technol. **5**(9), 2200059 (2022) [doi:10.1002/qute.202200059].

199.    D. Scognamiglio et al., "On-demand quantum light sources for underwater communications," Mater. Quantum Technol. **4**(2), 25402 (2024) [doi:10.1088/2633-4356/ad46d7].

200.    S. Vaidya et al., "Quantum sensing and imaging with spin defects in hexagonal boron nitride," Advances in Physics: X **8**(1), 2206049 (2023) [doi:10.1080/23746149.2023.2206049].

201.    V. M. Acosta et al., "Temperature dependence of the nitrogen-vacancy magnetic resonance in diamond," Phys. Rev. Lett. **104**(7), 70801 (2010) [doi:10.1103/PhysRevLett.104.070801].

202.    M. W. Doherty et al., "Temperature shifts of the resonances of the NV − center in diamond," Phys. Rev. B **90**(4), 41201 (2014) [doi:10.1103/PhysRevB.90.041201].

203.    A. Gottscholl et al., "Initialization and read-out of intrinsic spin defects in a van der Waals crystal at room temperature," Nat. Mater. **19**(5), 540–545 (2020) [doi:10.1038/s41563-020-0619-6].

204.    W. Liu et al., "Temperature-Dependent Energy-Level Shifts of Spin Defects in Hexagonal Boron Nitride," ACS Photonics **8**(7), 1889–1895 (2021) [doi:10.1021/acsphotonics.1c00320].

205.    N. Mathur et al., "Excited-state spin-resonance spectroscopy of $V_{B}^{-}$ defect centers in hexagonal boron nitride," Nat Commun **13**(1), 3233 (2022) [doi:10.1038/s41467-022-30772-z].

206.    A. Gottscholl et al., "Spin defects in hBN as promising temperature, pressure and magnetic field quantum sensors," Nat Commun **12**(1), 4480 (2021) [doi:10.1038/s41467-021-24725-1].

207.    M. Huang et al., "Wide field imaging of van der Waals ferromagnet Fe3GeTe2 by spin defects in hexagonal boron nitride," Nat Commun **13**(1), 5369 (2022) [doi:10.1038/s41467-022-33016-2].

208.    X. Gao et al., "Quantum Sensing of Paramagnetic Spins in Liquids with Spin Qubits in Hexagonal Boron Nitride," ACS Photonics **10**(8), 2894–2900 (2023) [doi:10.1021/acsphotonics.3c00621].

209.     B. C. Cavenett, "Optically detected magnetic resonance (O.D.M.R.) investigations of recombination processes in semiconductors," Adv. Phys. **30**(4), 475–538 (1981) [doi:10.1080/00018738100101397].

210.     P. Yu et al., "Excited-State Spectroscopy of Spin Defects in Hexagonal Boron Nitride," Nano Lett. **22**(9), 3545–3549 (2022) [doi:10.1021/acs.nanolett.1c04841].

211.     P. Khatri et al., "Optical Gating of Photoluminescence from Color Centers in Hexagonal Boron Nitride," Nano Lett. **20**(6), 4256–4263 (2020) [doi:10.1021/acs.nanolett.0c00751].

212.     N. Mendelson et al., "Identifying carbon as the source of visible single-photon emission from hexagonal boron nitride," Nat. Mater. **20**(3), 321–328 (2021) [doi:10.1038/s41563-020-00850-y].

213.     A. Dietrich et al., "Solid-state single photon source with Fourier transform limited lines at room temperature," Phys. Rev. B **101**(8), 081401 (2020) [doi:10.1103/PhysRevB.101.081401].

214.     J. E. Fröch et al., "Coupling Hexagonal Boron Nitride Quantum Emitters to Photonic Crystal Cavities," ACS Nano **14**(6), 7085–7091 (2020) [doi:10.1021/acsnano.0c01818].

215.     H. L. Stern et al., "Room-temperature optically detected magnetic resonance of single defects in hexagonal boron nitride," Nat Commun **13**(1), 618 (2022) [doi:10.1038/s41467-022-28169-z].

216.     N.-J. Guo et al., "Coherent control of an ultrabright single spin in hexagonal boron nitride at room temperature," Nat Commun **14**(1), 2893 (2023) [doi:10.1038/s41467-023-38672-6].

217.     H. L. Stern et al., "A quantum coherent spin in hexagonal boron nitride at ambient conditions," Nat. Mater. **23**(10), 1379–1385 (2024) [doi:10.1038/s41563-024-01887-z].

218.     N. Chejanovsky et al., "Single-spin resonance in a van der waals embedded paramagnetic defect," Nat. Mater. **20**(8), 1079–1084 (2021) [doi:10.1038/s41563-021-00979-4].

219.     S. C. Scholten et al., "Multi-species optically addressable spin defects in a van der Waals material," Nat Commun **15**(1), 6727 (2024) [doi:10.1038/s41467-024-51129-8].

220.     A. J. Healey et al., "Quantum microscopy with van der Waals heterostructures," Nat. Phys. **19**(1), 87–91 (2023) [doi:10.1038/s41567-022-01815-5].

221.    K. Sasaki et al., "Magnetic field imaging by hBN quantum sensor nanoarray," Appl. Phys. Lett. **122**(24), 244003 (2023) [doi:10.1063/5.0147072].

222.    J. S. Moon et al., "Fiber‑Integrated van der Waals Quantum Sensor with an Optimal Cavity Interface," Advanced Optical Materials, 2401987 (2024) [doi:10.1002/adom.202401987].

223.    T. Kosmala et al., "Strain induced phase transition of WS2 by local dewetting of Au/mica film upon annealing," Surfaces **4**(1), 1–8 (2020) [doi:10.3390/surfaces4010001].

224.    K. F. Mak et al., "Atomically Thin MoS 2 : A New Direct-Gap Semiconductor," Phys. Rev. Lett. **105**(13), 136805 (2010) [doi:10.1103/PhysRevLett.105.136805].

225.    A. Splendiani et al., "Emerging photoluminescence in monolayer MoS $_2$," Nano Lett. **10**(4), 1271–1275 (2010) [doi:10.1021/nl903868w].

226.    B. Zhu, X. Chen, and X. Cui, "Exciton binding energy of monolayer WS2," Sci. Rep. **5**(1), 9218 (2015) [doi:10.1038/srep09218].

227.    S. Park et al., "Direct determination of monolayer MoS $_2$ and WSe $_2$ exciton binding energies on insulating and metallic substrates," 2D Mater. **5**(2), 25003 (2018) [doi:10.1088/2053-1583/aaa4ca].

228.    X. Lu et al., "Optical initialization of a single spin-valley in charged WSe2 quantum dots," Nat. Nanotechnol. **14**(5), 426–431 (2019) [doi:10.1038/s41565-019-0394-1].

229.    J. Kern et al., "Nanoscale Positioning of Single‑Photon Emitters in Atomically Thin WSe $_2$," Advanced Materials **28**(33), 7101–7105 (2016) [doi:10.1002/adma.201600560].

230.    S. Zhang et al., "Defect structure of localized excitons in a WSe 2 monolayer," Phys. Rev. Lett. **119**(4), 46101 (2017) [doi:10.1103/PhysRevLett.119.046101].

231.    G. D. Shepard et al., "Nanobubble induced formation of quantum emitters in monolayer semiconductors," 2D Mater. **4**(2), 021019 (2017) [doi:10.1088/2053-1583/aa629d].

232.    S. Cianci et al., "Spatially Controlled Single Photon Emitters in hBN‑Capped WS $_2$ Domes," Advanced Optical Materials **11**(12), 2202953 (2023) [doi:10.1002/adom.202202953].

233.    C. Palacios-Berraquero et al., "Large-scale quantum-emitter arrays in atomically thin semiconductors," Nat Commun **8**(1), 15093 (2017) [doi:10.1038/ncomms15093].

234.    W. Wu et al., "Locally defined quantum emission from epitaxial few-layer tungsten diselenide," Applied Physics Letters **114**(21), 213102 (2019) [doi:10.1063/1.5091779].

235.    Y. Luo et al., "Deterministic coupling of site-controlled quantum emitters in monolayer WSe2 to plasmonic nanocavities," Nature Nanotech **13**(12), 1137–1142 (2018) [doi:10.1038/s41565-018-0275-z].

236.    M. R. Rosenberger et al., "Quantum Calligraphy: Writing Single-Photon Emitters in a Two-Dimensional Materials Platform," ACS Nano **13**(1), 904–912 (2019) [doi:10.1021/acsnano.8b08730].

237.    J.-P. So et al., "Electrically driven strain-induced deterministic single-photon emitters in a van der Waals heterostructure," Sci. Adv. **7**(43), eabj3176 (2021) [doi:10.1126/sciadv.abj3176].

238.    G. Moody et al., "Microsecond Valley Lifetime of Defect-Bound Excitons in Monolayer WSe 2," Phys. Rev. Lett. **121**(5), 057403 (2018) [doi:10.1103/PhysRevLett.121.057403].

239.    K. Parto et al., "Defect and strain engineering of monolayer WSe2 enables site-controlled single-photon emission up to 150 K," Nat Commun **12**(1), 3585 (2021) [doi:10.1038/s41467-021-23709-5].

240.    C. Palacios-Berraquero et al., "Atomically thin quantum light-emitting diodes," Nat Commun **7**(1), 12978 (2016) [doi:10.1038/ncomms12978].

241.    J. Klein et al., "Site-selectively generated photon emitters in monolayer MoS2 via local helium ion irradiation," Nat Commun **10**(1), 2755 (2019) [doi:10.1038/s41467-019-10632-z].

242.    K. Barthelmi et al., "Atomistic defects as single-photon emitters in atomically thin MoS2," Applied Physics Letters **117**(7), 070501 (2020) [doi:10.1063/5.0018557].

243.    J. Klein et al., "Engineering the luminescence and generation of individual defect emitters in atomically thin MoS 2," ACS Photonics **8**(2), 669–677 (2021) [doi:10.1021/acsphotonics.0c01907].

244.    A. Branny et al., "Discrete quantum dot like emitters in monolayer MoSe2: Spatial mapping, magneto-optics, and charge tuning," Applied Physics Letters **108**(14), 142101 (2016) [doi:10.1063/1.4945268].

245.    L. Yu et al., "Site-controlled quantum emitters in monolayer MoSe 2," Nano Lett. **21**(6), 2376–2381 (2021) [doi:10.1021/acs.nanolett.0c04282].

246.    H. Zhao et al., "Manipulating Interlayer Excitons for Near-Infrared Quantum Light Generation," Nano Lett. **23**(23), 11006–11012 (2023) [doi:10.1021/acs.nanolett.3c03296].

247.   F. He et al., "Moiré patterns in 2D materials: a review," ACS Nano **15**(4), 5944–5958 (2021) [doi:10.1021/acsnano.0c10435].

248.   K. L. Seyler et al., "Signatures of moiré-trapped valley excitons in MoSe2/WSe2 heterobilayers," Nature **567**(7746), 66–70 (2019) [doi:10.1038/s41586-019-0957-1].

249.   H. Baek et al., "Highly energy-tunable quantum light from moiré-trapped excitons," Sci. Adv. **6**(37), eaba8526 (2020) [doi:10.1126/sciadv.aba8526].

250.   H.-J. Chuang et al., "Enhancing single photon emission purity via design of van der waals heterostructures," Nano Lett. **24**(18), 5529–5535 (2024) [doi:10.1021/acs.nanolett.4c00683].

251.   L. Sortino et al., "Bright single photon emitters with enhanced quantum efficiency in a two-dimensional semiconductor coupled with dielectric nano-antennas," Nat Commun **12**(1), 6063 (2021) [doi:10.1038/s41467-021-26262-3].

252.   M. Von Helversen et al., "Temperature dependent temporal coherence of metallic-nanoparticle-induced single-photon emitters in a WSe$_2$ monolayer," 2D Mater. **10**(4), 045034 (2023) [doi:10.1088/2053-1583/acfb20].

253.   J.-C. Drawer et al., "Monolayer-Based Single-Photon Source in a Liquid-Helium-Free Open Cavity Featuring 65% Brightness and Quantum Coherence," Nano Lett. **23**(18), 8683–8689 (2023) [doi:10.1021/acs.nanolett.3c02584].

254.   T. Gao et al., "Atomically-thin single-photon sources for quantum communication," npj 2D Mater Appl **7**(1), 4 (2023) [doi:10.1038/s41699-023-00366-4].

255.   G. Grosso et al., "Tunable and high-purity room temperature single-photon emission from atomic defects in hexagonal boron nitride," Nat Commun **8**(1), 705 (2017) [doi:10.1038/s41467-017-00810-2].

256.   Y. Xue et al., "Anomalous Pressure Characteristics of Defects in Hexagonal Boron Nitride Flakes," ACS Nano **12**(7), 7127–7133 (2018) [doi:10.1021/acsnano.8b02970].

257.   O. Iff et al., "Strain-tunable single photon sources in WSe$_2$ monolayers," Nano Lett. **19**(10), 6931–6936 (2019) [doi:10.1021/acs.nanolett.9b02221].

258.   G. Noh et al., "Stark Tuning of Single-Photon Emitters in Hexagonal Boron Nitride," Nano Lett. **18**(8), 4710–4715 (2018) [doi:10.1021/acs.nanolett.8b01030].

259.   C. Chakraborty et al., "Quantum-Confined Stark Effect of Individual Defects in a van der Waals Heterostructure," Nano Lett. **17**(4), 2253–2258 (2017) [doi:10.1021/acs.nanolett.6b04889].

260.    N. Nikolay et al., "Very large and reversible stark-shift tuning of single emitters in layered hexagonal boron nitride," Phys. Rev. Appl. **11**(4), 41001 (2019) [doi:10.1103/PhysRevApplied.11.041001].

261.    I. Zhigulin et al., "Stark effect of blue quantum emitters in hexagonal boron nitride," Phys. Rev. Appl. **19**(4), 44011 (2023) [doi:10.1103/PhysRevApplied.19.044011].

262.    A. Ciarrocchi et al., "Polarization switching and electrical control of interlayer excitons in two-dimensional van der waals heterostructures," Nat. Photonics **13**(2), 131–136 (2019) [doi:10.1038/s41566-018-0325-y].

263.    M. I. B. Utama et al., "Chemomechanical modification of quantum emission in monolayer WSe2," Nat Commun **14**(1), 2193 (2023) [doi:10.1038/s41467-023-37892-0].

264.    E. T. Jaynes and F. W. Cummings, "Comparison of quantum and semiclassical radiation theories with application to the beam maser," Proc. IEEE **51**(1), 89–109 (1963) [doi:10.1109/PROC.1963.1664].

265.    T. Grange et al., "Cavity-funneled generation of indistinguishable single photons from strongly dissipative quantum emitters," Phys. Rev. Lett. **114**(19), 193601 (2015) [doi:10.1103/PhysRevLett.114.193601].

266.    C. Zhang et al., "Microstructure Engineering of Hexagonal Boron Nitride for Single‐Photon Emitter Applications," Advanced Optical Materials **10**(17), 2200207 (2022) [doi:10.1002/adom.202200207].

267.    J.-H. Kim et al., "Two-photon interference from a bright single-photon source at telecom wavelengths," Optica **3**(6), 577 (2016) [doi:10.1364/OPTICA.3.000577].

268.    X. Ding et al., "High-efficiency single-photon source above the loss-tolerant threshold for efficient linear optical quantum computing," arXiv:2311.08347, arXiv (2023).

269.    A. Jeantet et al., "Widely tunable single-photon source from a carbon nanotube in the purcell regime," Phys. Rev. Lett. **116**(24), 247402 (2016) [doi:10.1103/PhysRevLett.116.247402].

270.    A. Jeantet et al., "Exploiting one-dimensional exciton–phonon coupling for tunable and efficient single-photon generation with a carbon nanotube," Nano Lett. **17**(7), 4184–4188 (2017) [doi:10.1021/acs.nanolett.7b00973].

271.    D. Hunger et al., "A fiber Fabry–Perot cavity with high finesse," New J. Phys. **12**(6), 065038 (2010) [doi:10.1088/1367-2630/12/6/065038].

272.    T. Grange et al., "Cavity-Funneled Generation of Indistinguishable Single Photons from Strongly Dissipative Quantum Emitters," Phys. Rev. Lett. **114**(19), 193601 (2015) [doi:10.1103/PhysRevLett.114.193601].

273.    T. T. Tran et al., "Deterministic coupling of quantum emitters in 2D materials to plasmonic nanocavity arrays," Nano Lett. **17**(4), 2634–2639 (2017) [doi:10.1021/acs.nanolett.7b00444].

274.    T. Cai et al., "Coupling emission from single localized defects in two-dimensional semiconductor to surface plasmon polaritons," Nano Lett. **17**(11), 6564–6568 (2017) [doi:10.1021/acs.nanolett.7b02222].

275.    L. C. Flatten et al., "Microcavity enhanced single photon emission from two-dimensional WSe2," Appl. Phys. Lett. **112**(19), 191105 (2018) [doi:10.1063/1.5026779].

276.    C. Errando-Herranz et al., "Resonance fluorescence from waveguide-coupled, strain-localized, two-dimensional quantum emitters," ACS Photonics **8**(4), 1069–1076 (2021) [doi:10.1021/acsphotonics.0c01653].

277.    Y.-C. Chen et al., "Laser writing of individual nitrogen-vacancy defects in diamond with near-unity yield," Optica **6**(5), 662 (2019) [doi:10.1364/OPTICA.6.000662].

278.    N. P. Wilson et al., "Excitons and emergent quantum phenomena in stacked 2D semiconductors," Nature **599**(7885), 383–392 (2021) [doi:10.1038/s41586-021-03979-1].

279.    D. A. Bandurin et al., "High electron mobility, quantum Hall effect and anomalous optical response in atomically thin InSe," Nature Nanotech **12**(3), 223–227 (2017) [doi:10.1038/nnano.2016.242].

280.    T. V. Shubina et al., "InSe as a case between 3D and 2D layered crystals for excitons," Nat. Commun. **10**(1), 3479 (2019) [doi:10.1038/s41467-019-11487-0].

281.    M. Salomone et al., "Point defects in two-dimensional indium selenide as tunable single-photon sources," J. Phys. Chem. Lett. **12**(45), 10947–10952 (2021) [doi:10.1021/acs.jpclett.1c02912].

282.    W. Luo et al., "Deterministic Localization of Strain-Induced Single-Photon Emitters in Multilayer GaSe," ACS Photonics **10**(8), 2530–2539 (2023) [doi:10.1021/acsphotonics.3c00052].

283.    W. Luo et al., "Improving Strain-localized GaSe Single Photon Emitters with Electrical Doping," Nano Lett. **23**(21), 9740–9747 (2023) [doi:10.1021/acs.nanolett.3c02308].

284.    D. G. Hopkinson et al., "Formation and healing of defects in atomically thin GaSe and InSe," ACS Nano **13**(5), 5112–5123 (2019) [doi:10.1021/acsnano.8b08253].

285.    A. S. Sarkar and E. Stratakis, "Recent advances in 2D metal monochalcogenides," Adv. Sci. **7**(21), 2001655 (2020) [doi:10.1002/advs.202001655].

286.    Z. Hu et al., "Recent progress in 2D group IV–IV monochalcogenides: synthesis, properties and applications," Proc. Spie. **30**(25), 252001 (2019) [doi:10.1088/1361-6528/ab07d9].

287.    G. W. Mudd et al., "The direct-to-indirect band gap crossover in two-dimensional van der waals indium selenide crystals," Sci. Rep. **6**(1), 39619 (2016) [doi:10.1038/srep39619].

288.    P. Tonndorf et al., "On-Chip Waveguide Coupling of a Layered Semiconductor Single-Photon Source," Nano Lett. **17**(9), 5446–5451 (2017) [doi:10.1021/acs.nanolett.7b02092].

289.    D. Jariwala, T. J. Marks, and M. C. Hersam, "Mixed-dimensional van der Waals heterostructures," Nature Mater **16**(2), 170–181 (2017) [doi:10.1038/nmat4703].

290.    O. Lopez-Sanchez et al., "Light generation and harvesting in a van der waals heterostructure," ACS Nano **8**(3), 3042–3048 (2014) [doi:10.1021/nn500480u].

291.    L. Dou et al., "Atomically thin two-dimensional organic-inorganic hybrid perovskites," Science **349**(6255), 1518–1521 (2015) [doi:10.1126/science.aac7660].

292.    X. Li et al., "Proximity-induced chiral quantum light generation in strain-engineered WSe2/NiPS3 heterostructures," Nat. Mater. **22**(11), 1311–1316 (2023) [doi:10.1038/s41563-023-01645-7].

293.    K. Azuma et al., "Quantum repeaters: From quantum networks to the quantum internet," Rev. Mod. Phys. **95**(4), 045006 (2023) [doi:10.1103/RevModPhys.95.045006].

294.    F. Ewert and P. van Loock, "$3/4$-Efficient Bell Measurement with Passive Linear Optics and Unentangled Ancillae," Phys. Rev. Lett. **113**(14), 140403 (2014) [doi:10.1103/PhysRevLett.113.140403].

295.    E. Kaur, A. Patil, and S. Guha, "Resource-efficient and loss-aware photonic graph state preparation using an array of quantum emitters, and application to all-photonic quantum repeaters," arXiv.org, 1 February 2024, <https://arxiv.org/abs/2402.00731v1> (accessed 14 October 2024).

296.    A. Reiserer and G. Rempe, "Cavity-based quantum networks with single atoms and optical photons," Rev. Mod. Phys. **87**(4), 1379–1418 (2015) [doi:10.1103/RevModPhys.87.1379].

297.    S. Sun et al., "A quantum phase switch between a single solid-state spin and a photon," Nature Nanotechnology **11**(6), 539–544 (2016) [doi:10.1038/nnano.2015.334].

298.    M. K. Bhaskar et al., "Experimental demonstration of memory-enhanced quantum communication," Nature **580**(7801), 60–64 (2020) [doi:10.1038/s41586-020-2103-5].

299.    A. Reiserer, "Colloquium: Cavity-enhanced quantum network nodes," Rev. Mod. Phys. **94**(4), 041003 (2022) [doi:10.1103/RevModPhys.94.041003].

300.    J. Gao et al., "Selective wrapping and supramolecular structures of polyfluorene–carbon nanotube hybrids," ACS Nano **5**(5), 3993–3999 (2011) [doi:10.1021/nn200564n].

301.    J. Kern et al., "Nanoscale Positioning of Single‐Photon Emitters in Atomically Thin $WSe_2$," Advanced Materials **28**(33), 7101–7105 (2016) [doi:10.1002/adma.201600560].